\documentclass[10pt,prb,aps,superscriptaddress,eqsecnum,floatfix,twocolumn,showpacs,amsfonts,amsmath,amssymb,final,superscriptaddress]{revtex4-1}

\usepackage{times}
\usepackage{graphicx}
\usepackage{stmaryrd}
\usepackage{rotating}
\usepackage{wasysym}
\usepackage{amsmath}
\usepackage{amsfonts}
\usepackage{amssymb}
\usepackage{braket}
\usepackage[countmax]{subfloat}
\usepackage{dcolumn}
\usepackage{bm}
\usepackage{color}

\def\:={\,\raisebox{0.85pt}{.}\hspace{-2.78pt}\raisebox{2.85pt}{.}\!\!=\,}
\def\=:{\,=\!\!\raisebox{0.85pt}{.}\hspace{-2.78pt}\raisebox{2.85pt}{.}\,}

\usepackage{float}
\usepackage{latexsym}
\usepackage[breaklinks]{hyperref}
\hypersetup{colorlinks=true, linkcolor=blue, citecolor=blue, filecolor=blue, urlcolor=blue}
\usepackage{cancel,soul}
\usepackage[normalem]{ulem}

\begin{document}

\title{
Weyl-type topological phase transitions in fractional quantum Hall like systems
      }

\author{Stefanos Kourtis}

\affiliation{Department of Physics, Boston University, Boston, MA, 02215, USA}

\author{Titus Neupert}

\affiliation{Department of Physics, University of Zurich, Winterthurerstrasse
190, CH-8057 Zurich, Switzerland}

\author{Christopher Mudry}

\affiliation{Condensed Matter Theory Group, Paul Scherrer Institute,
CH-5232 Villigen PSI, Switzerland}

\author{Manfred Sigrist}

\affiliation{Institute for Theoretical Physics, ETH Zurich, CH-8093 Zurich,
Switzerland}

\author{Wei Chen}

\affiliation{Institute for Theoretical Physics, ETH Zurich, CH-8093 Zurich,
Switzerland}

\date{\rm\today}

\begin{abstract}

We develop a method to characterize topological phase transitions for
strongly correlated Hamiltonians defined on two-dimensional lattices
based on the many-body Berry curvature. Our goal is to identify a
class of quantum critical points between topologically nontrivial
phases with fractionally quantized Hall (FQH) conductivity and
topologically trivial gapped phases through the discontinuities of the
many-body Berry curvature in the so-called flux Brillouin zone (fBZ),
the latter being defined by imposing all possible twisted boundary
conditions. For this purpose, we study the finite-size signatures of
several quantum phase transitions between fractional Chern insulators
and charge-ordered phases for two-dimensional lattices by evaluating
the many-body Berry curvature numerically using exact
diagonalization. We observe degeneracy points (nodes) of many-body
energy levels at high-symmetry points in the fBZ, accompanied by
diverging Berry curvature. We find a correspondence between the number
and order of these nodal points, and the change of the topological
invariants of the many-body ground states across the transition, in
close analogy with Weyl nodes in non-interacting band structures. This
motivates us to apply a scaling procedure, originally developed for
non-interacting systems, for the Berry curvature at the nodal
points. This procedure offers a useful tool for the classification of
topological phase transitions in interacting systems harboring
FQH-like topological order.

\end{abstract}

\maketitle

\section{Introduction} 
\label{sec: Introduction}

Degeneracies in quantum systems can be accidental or enforced by
symmetries. In the former case, slight perturbations applied to the
system can in principle remove the degeneracy. Choosing words more
carefully, one should distinguish between \emph{moving} and
\emph{removing} a degeneracy. To elucidate this point, we consider a
pair of linearly independent quantum states
$|\Psi^{\,}_{1}(\boldsymbol{M})\rangle$ and
$|\Psi^{\,}_{2}(\boldsymbol{M})\rangle$ that are parametrized by a
multi-dimensional vector of real parameters $\boldsymbol{M}$.
The matrix elements of a given Hamiltonian with the two states
$|\Psi^{\,}_{1}(\boldsymbol{M})\rangle$ and
$|\Psi^{\,}_{2}(\boldsymbol{M})\rangle$
form a $2\times2$ Hermitian matrix, and are thus characterized by
four real functions of $\boldsymbol{M}$,
e.g., as prefactors when expanding in the three Pauli matrices
together with the unit $2\times2$ matrix. The states are degenerate if
all three functions multiplying the three Pauli matrices
are tuned to zero at some
$\boldsymbol{M}^{\,}_{\mathrm{c}}$.  Such simultaneous zeros will occur
generically, if the parameter space in which $\boldsymbol{M}$ lives is
three-dimensional, in which case perturbations applied to
such accidental degeneracies do not remove them but move them
in parameter space. If the dimension of parameter space $d$ is smaller
than three, three independent real functions cannot be simultaneously
tuned to zero, while if it is larger than three,
accidental degeneracies will occur on $(d-3)$-dimensional hypersurfaces.
These considerations go back to
von Neumann and Wigner\cite{vonNeumann93}, while
Herring~\cite{Herring37b} and Blount~\cite{Blount62} applied this
reasoning to band structures of three-dimensional crystals, in which
the quasi-momentum $\boldsymbol{k}$ takes the role of the parameter
vector $\boldsymbol{M}$. Berry and Wilkinson also found
in Ref.\ \onlinecite{Wilkinson1984}
such degeneracies for quantum single-particle Hamiltonians without any
symmetry. They called such  degeneracies diabolical points.

Crucially, these accidental degeneracies found in three-dimensional
parameter space are topological objects with a quantized chiral
charge, so-called Weyl points. The terminology ``Weyl point''
can be motivated as follows. Consider the
two-dimensional Hilbert space spanned by the state
$|\Psi^{\,}_{1}(\boldsymbol{M})\rangle$
and
$|\Psi^{\,}_{2}(\boldsymbol{M})\rangle$
whereby both states undergo a level crossing
at a diabolical point in parameter space.
We may associate to any point in parameter space
that is sufficiently close to the diabolical point
an Abelian gauge field, the Abelian Berry connection
$
\langle\Psi^{\,}_{1}(\boldsymbol{M})|
\mathrm{i}\partial^{\,}_{\boldsymbol{M}}
|\Psi^{\,}_{1}(\boldsymbol{M})\rangle
$
of state
$|\Psi^{\,}_{1}(\boldsymbol{M})\rangle$, say.
The rotation of this Abelian Berry connection delivers an
Abelian magnetic field. The Abelian magnetic flux
through any surface that encloses the diabolical point is quantized
and thus robust to small and smooth changes in the Hamiltonian
~\cite{Volovikbook}. 
Any such diabolical point is therefore a
topological object with a quantized charge, i.e., a monopole for an Abelian
Berry magnetic flux. Because the four matrix elements of the Hamiltonian
between these two states realize a Weyl Hamiltonian close to a diabolical
point, we shall rename this point in parameter space a \textit{Weyl point}.
Generic perturbations will simply move a Weyl point
in parameter space, but cannot eliminate it.

In this paper, we study
numerically away from the thermodynamic limit Weyl points
associated with degeneracies between
two many-body quantum states close to a topological phase transition
occuring in parameter space in the thermodynamic limit.
Away from multicritical points, phase transitions are driven by a single
external parameter (i.e., the co-dimension of the phase boundary in
parameter space is one). It is then most natural to look for phase
transitions characterized by accidental Weyl-type degeneracies of two
quantum states in two-dimensional systems, because the phase-angles of
twisted boundary conditions comprise two parameters that add up with
a control parameter in the many-body interacting Hamiltonian
to a three-dimensional parameter space.
In this case, the topological charge associated with the degeneracy is
nothing but the \textit{change} 
in the Hall conductivity averaged over twisted boundary conditions
following the formula of Niu and Thouless
\cite{Niu1985a,fukui05}. This change has a finite-size signature
as it is unambiguously determined
by a single many-body level crossing for the many-body interacting
lattice Hamiltonian, \textit{whereby the lattice is of finite size}.
The main result of this paper is that,
in close analogy to noninteracting band structures over a
three-dimensional Brillouin zone,
the change in the Hall conductivity of any two-dimensional many-body
interacting lattice Hamiltonian subject to twisted boundary conditions
across a quantum phase transition controlled by one real parameter
equals the topological charge associated with a many-body Weyl point
between two many-body eigenstates.

Weyl-type quantum phase transitions are thus necessarily topological quantum
phase transitions and one may expect them to be realized in systems
that support, in one phase, a (fractional) quantum-Hall effect. This
is also in agreement with the fact that quantum Hall systems do not
rely on any symmetries (besides charge conservation), just as a Weyl
point of unit chiral charge is not stabilized by any symmetries.

Our approach to Weyl-type quantum phase transitions is focused on
numerical investigation of strongly interacting finite-sized
systems. To easily apply twisted boundary conditions, we study lattice
systems, and in particular the transition between lattice realizations
of the fractional quantum Hall effect, that is, fractional Chern
insulators (FCIs)~\cite{Neupert2011,Regnault2011},
and charge ordered phases. We demonstrate that these
quantum phase transitions are indeed of Weyl type.  We show examples
for both bosonic and fermionic systems and present a case in which
lattice symmetries give rise to a Weyl point of higher charge (two) at
the phase transition, which splits into two Weyl points of unit charge if these symmetries are broken.
Furthermore, we point out the similarities and
differences between the physics of Weyl points occurring between
many-body states and those in non-interacting band structures, and
apply a scaling procedure that captures the numerically observed
phenomenology of the Weyl-type quantum phase transitions.

This paper is organized as follows.
We review the Niu-Thouless formula in Sec.\
\ref{sec: Berry curvature and Chern number ...}.
We define lattice models for interacting fermions or bosons that
undergo a quantum phase transition between a fractional Chern insulating phase
and a trivial one (i.e., a phase with vanishing Hall conductivity)
and show numerically the existence of
Weyl (diabolical) points
in parameter space in Sec.\ \ref{sec: FCI-to-trivial transitions}.
We verify that a scaling analysis for the many-body Berry curvature
applies in Sec.\ \ref{sec:BCRG}.
We conclude with Sec.\ \ref{sec: Summary and conclusion}.

\section{Berry curvature and Chern number of many-body states\label{sec:Berry_curvature_Chern_number}}
\label{sec: Berry curvature and Chern number ...}

Consider a two-dimensional system of $N$ interacting
identical quantum particles
on a lattice made of
$L^{\,}_{1} \times L^{\,}_{2}$
unit cells, defined by the primitive translation vectors
$\bm{a}^{\,}_{1}$ and $\bm{a}^{\,}_{2}$. The system is described by a
many-body Hamiltonian $\widehat{H}(\bm{M})$
that depends on a number of parameters with units of energy 
$\bm{M}\equiv(M^{\,}_{1},\cdots,M^{\,}_{m})^{\mathsf{T}}\in\mathbb{R}^{m}$. 
We also impose twisted periodic boundary conditions on the system.
This amounts to introducing a second parametric dependence
of the Hamiltonian on the twisting angles given by the vector
$\bm{\phi}^{\mathsf{T}}=(\phi^{\,}_{1},\phi^{\,}_{2})$, i.e.,
$\widehat{H}\equiv\widehat{H}(\bm{\phi},\bm{M})$. Hence, all many-body states
$|\Psi(\bm{\phi},\bm{M})\rangle$
obey the twisted boundary conditions
\begin{subequations}
\label{eq:twisted_bc}
\begin{align}
\begin{split}
&
\widehat{T}^{\,}_{i,L^{\,}_{1}\,\bm{a}^{\,}_{1}}
|\Psi(\bm{\phi},\bm{M})\rangle=
e^{\mathrm{i}\phi^{\,}_{1}}\,
|
\Psi(\bm{\phi},\bm{M})\rangle,
\end{split}
\\
\begin{split}
&
\widehat{T}^{\,}_{i,L^{\,}_{2}\,\bm{a}^{\,}_{2}}
|\Psi(\bm{\phi},\bm{M})\rangle=
e^{\mathrm{i}\phi^{\,}_{2}}\,
|
\Psi(\bm{\phi},\bm{M})\rangle,
\end{split}
\end{align}
\end{subequations}
where $\widehat{T}^{\,}_{i,\bm{r}}$ is the operator that translates particle
$i=1,\cdots,N$ by $\bm{r}$. 

The Hall conductivity $\sigma^{\,}_{\mathrm{H}}(\bm{\phi},\bm{M})$
at zero temperature governs the transverse linear response to
the infinitesimal variation
$\bm{\phi}\to\bm{\phi}+\delta\bm{\phi}$. 
It is given by~\cite{Thouless1982,Niu1985a}
\begin{subequations}
\begin{equation}
\sigma^{\,}_{\mathrm{H}}(\bm{\phi},\bm{M})=
\frac{e^{2}}{h}
\frac{1}{N^{\,}_{g=1}}
\sum_{n=1}^{N^{\,}_{g=1}}
F^{\,}_{n}(\bm{\phi},\bm{M}).
\label{eq:hall}
\end{equation}
Here, $e$ is the electric charge, $h$ is the Planck constant, and
the integer $N^{\,}_{g=1}$
is the ground state (GS) degeneracy in the thermodynamic limit
upon imposing the twisted boundary conditions $\bm{\phi}$. Moreover,
the function
$F^{\,}_{n}(\bm{\phi},\bm{M})$
is usually defined through linear response theory by the Kubo formula
\begin{equation}
F^{\,}_{n}(\bm{\phi},\bm{M})\:=
4\pi
\mathrm{Im}\!
\sum_{n'\not=n}\!
\frac{\bra{\Psi^{\,}_{n}}
\partial^{\,}_{\phi^{\,}_{2}}
\widehat{H}
\ket{\Psi^{\,}_{n'}}
\bra{\Psi^{\,}_{n'}}
\partial^{\,}_{\phi^{\,}_{1}}
\widehat{H}
\ket{\Psi^{\,}_{n}}}{(E^{\,}_{n'}-E^{\,}_{n})^{2}}
\label{eq:berry}
\end{equation}
provided all denominators on the right-hand side are non-vanishing in a
sufficiently small open set containing $\bm{\phi}$ and $\bm{M}$.
Here, the summation over $n'$ runs over all the eigenstates 
$\ket{\Psi^{\,}_{n'}(\bm{\phi},\bm{M})}$ of
$\widehat{H}(\bm{\phi},\bm{M})$
that are orthogonal to any one of the GSs
$\ket{\Psi^{\,}_{n}(\bm{\phi},\bm{M})}$ 
with $n=1,\cdots,N^{\,}_{g=1}$.
Their eigenenergies are $E^{\,}_{n'}(\bm{\phi},\bm{M})$
and $E^{\,}_{n}(\bm{\phi},\bm{M})$,
respectively. In the thermodynamic limit and on the torus (genus $g=1$)
over which the twisted boundary conditions are imposed,
the eigenenergies $E^{\,}_{n}(\bm{\phi},\bm{M})$
with $n=1,\cdots,N^{\,}_{g=1}$
are all degenerate. For notational simplicity, we have
dropped the explicit dependence of the Hamiltonian with its eigenenergies
and eigenstates on $\bm{\phi}$ and $\bm{M}$
in Eq.\ (\ref{eq:berry}).
The Hall conductivity (\ref{eq:hall})
is averaged over all states in the manifold of
$N^{\,}_{g=1}$-fold degenerate GS when two-dimensional
space is the twisted torus and the
thermodynamic limit is defined by taking the limit
$L^{\,}_1,L^{\,}_2\to\infty$
while holding the particle density $N/(L^{\,}_1 \, L^{\,}_2)$ fixed.

The Hall conductivity at zero temperature should only depend on the
linearly independent GSs of the many-body interacting Hamiltonian
$\widehat{H}$. This fact is not explicit in Eqs.\
(\ref{eq:hall}) and (\ref{eq:berry})
as all many-body excited states and their eigenvalues
contribute to the Hall conductivity. It turns out that
the quantity $F^{\,}_{n}(\bm{\phi},\bm{M})$ is
called the many-body Berry curvature of
$\ket{\Psi^{\,}_{n}(\bm{\phi},\bm{M})}$
as it can be expressed in terms of the  partial derivatives of
$\ket{\Psi^{\,}_{n}(\bm{\phi},\bm{M})}$
according to the formula~\cite{Thouless1982,Niu1985a}
\begin{equation}
F^{\,}_{n}(\bm{\phi},\bm{M})=
4\pi\,
\mathrm{Im}
\left\langle\left.
\frac{
\partial\Psi^{\,}_{n}
     }
     {
\partial\phi^{\,}_{1}
     }
(\bm{\phi},\bm{M})
\right|
\frac{
\partial\Psi^{\,}_{n}
     }
     {
\partial\phi^{\,}_{2}
     }
 (\bm{\phi},\bm{M})
\right\rangle
\label{eq:berry-alt}
\end{equation}
\end{subequations}
provided all denominators in Eq.\
(\ref{eq:berry}) are non-vanishing in a
sufficiently small open set containing $\bm{\phi}$ and $\bm{M}$.
According to Eq.\ (\ref{eq:berry-alt}),
the Hall conductivity (\ref{eq:hall}) is now explicitly solely
dependent on the linearly independent GSs.

In Ref.~\onlinecite{Niu1985a}, it is argued that the
equality
\begin{equation}
\lim_{L^{\,}_1,L^{\,}_2\to\infty}
\sigma^{\,}_{\mathrm{H}}(\bm{\phi}^{\,}_{0},\bm{M})=
\lim_{L^{\,}_1,L^{\,}_2\to\infty}
\int\limits_{0}^{2\pi}
\int\limits_{0}^{2\pi}
\frac{\mathrm{d}\phi^{\,}_{1}\,\mathrm{d}\phi^{\,}_{2}}{4\pi^{2}}
\sigma^{\,}_{\mathrm{H}}(\bm{\phi},\bm{M})
\label{eq: Hall quantization}
\end{equation}
must necessarily hold for any arbitrarily chosen
twisted boundary condition $\bm{\phi}^{\,}_{0}$
in the thermodynamic limit.
Here, the
domain of integration is the torus defined by the twisted boundary
conditions, to which we shall refer as the flux Brillouin zone
(fBZ). The intuition for
Eq.\ \eqref{eq: Hall quantization}
is that the choice of the boundary conditions should not affect
the values taken by the components of the conductivity tensor
in an insulating phase after the thermodynamic limit has been taken.

\begin{figure}[t]
 \centering
\includegraphics[width=0.99\columnwidth]{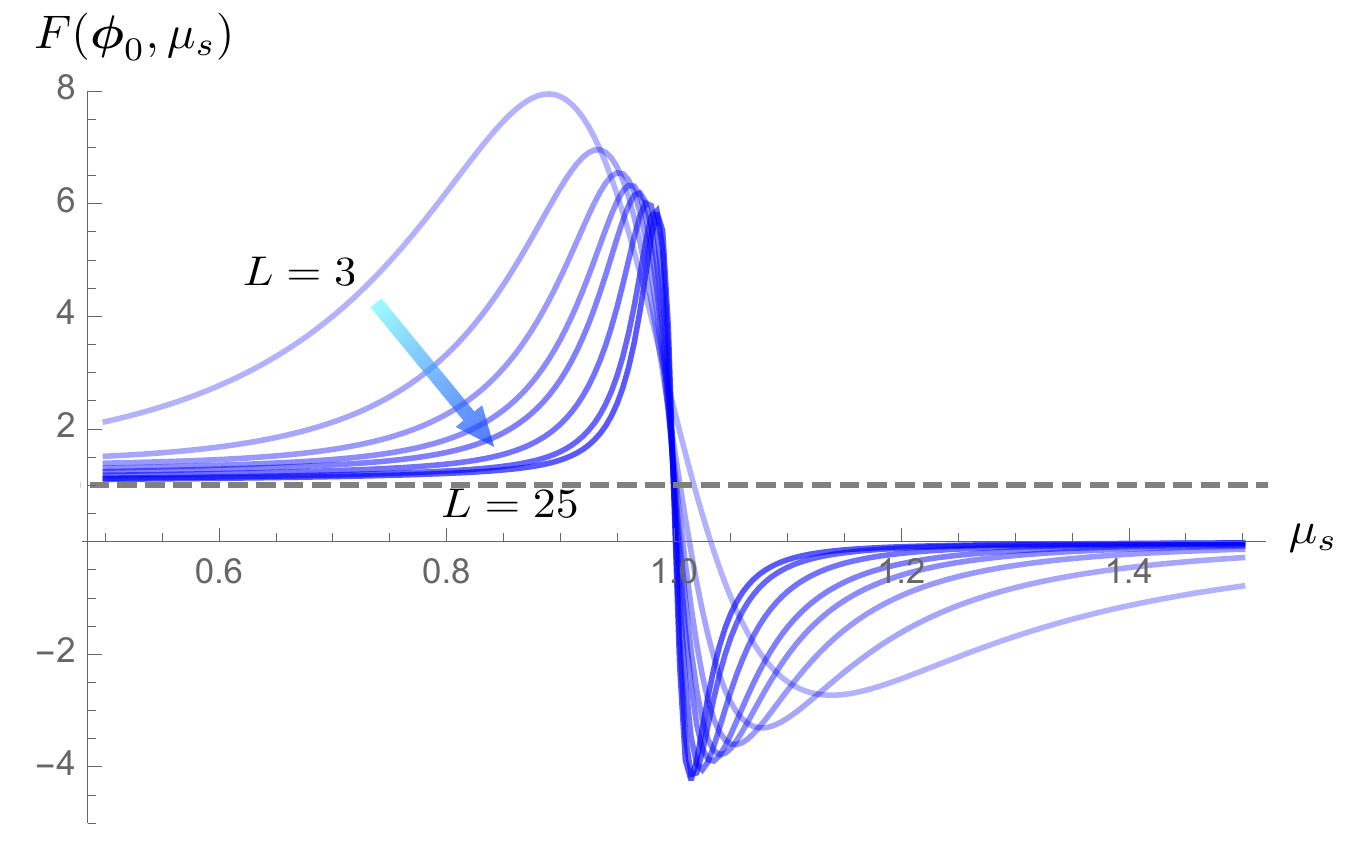}
\caption{
Many-body Berry curvature $F(\boldsymbol{\phi}^{\,}_{0},\mu^{\,}_{\mathrm{s}})$
for a non-interacting Chern insulator [specifically, the triangular
lattice model defined by the \textit{non-interacting} Hamiltonian
\eqref{eq:tri}] for fixed value
of flux $\boldsymbol{\phi}^{\,}_{0}=(\pi-0.7,0)$, for a series of system
sizes $L\equiv L^{\,}_{1}=L^{\,}_{2}=3, 5, 7, 9, 11, 15, 21, 25$. In the
thermodynamic limit, the model has a topological phase transition in
which the Chern number changes from 1 (gray line) to 0 at
$\mu^{\,}_{\mathrm{s}}=1$ (taking $t=1$ as energy unit).
In the thermodynamic limit,
$F(\boldsymbol{\phi}^{\,}_{0},\mu^{\,}_{\mathrm{s}})$ becomes constant and thus
independent of $\boldsymbol{\phi}^{\,}_{0}$ and $\mu^{\,}_{\mathrm{s}}$ except at
the phase transition. The many-body Berry curvature
$F(\boldsymbol{\phi},\mu^{\,}_{\mathrm{s}})$ is singular at
$\bm{\phi}^{\,}_{\mathrm{c}}=(\pi,0)$
and
$(\mu^{\,}_{\mathrm{s}}/t)^{\,}_{\mathrm{c}}=1$
for  $L\equiv L^{\,}_{1}=L^{\,}_{2}=3, 5, 7, 9, 11, 15, 21, 25$.}
\label{fig:MB Berry curvature}
\end{figure}

We note that Eq.\ \eqref{eq: Hall quantization} implies that,
as the thermodynamic limit is taken, the many-body Berry curvature
$F^{\,}_{n}(\bm{\phi},\bm{M})$ becomes quantized and thus independent
of $\bm{\phi}$ and $\bm{M}$, except for discontinuous changes at
$\bm{M}$ that correspond to a topological phase transition
at which the gap between the GS
manifold and the excited states closes. If so, at least one denominator
on the right-hand side of Eq.\ (\ref{eq:berry})
vanishes and the representation
(\ref{eq:berry-alt})
is ill defined.
We illustrate this for a non-interacting lattice model in Fig.\
\ref{fig:MB Berry curvature}. This is distinct from the Berry curvature of
single-particle bands, which generically has a momentum-dependence
even in the thermodynamic limit.  Hence, scaling relations obeyed by
$F(\boldsymbol{\phi},\bm{M})$ as a function of $\boldsymbol{\phi}$ and
$\bm{M}$, which we discuss in Sec.~\ref{sec:BCRG},
\textit{can only be defined in systems of finite size}.

Consequently, in the thermodynamic limit, 
\begin{subequations}
\begin{equation}
\frac{\sigma^{\,}_{\mathrm{H}}(\bm{M})}{(e^{2}/h)}\equiv
\frac{1}{N^{\,}_{g=1}}
\sum_{n=1}^{N^{\,}_{g=1}}
C^{\,}_{n}(\bm{M}),
\label{eq: Niu Thouless formula a}
\end{equation}
where
\begin{equation}
C^{\,}_{n}(\bm{M})\equiv
\lim_{L^{\,}_1,L^{\,}_2\to\infty}
\int\limits_{0}^{2\pi}\int\limits_{0}^{2\pi}
\frac{\mathrm{d}\phi^{\,}_{1}\,\mathrm{d}\phi^{\,}_{2}}{4\pi^2}
F^{\,}_{n}(\bm{\phi},\bm{M})
\label{eq: Niu Thouless formula b}
\end{equation}
\end{subequations}
is an integer-valued function of $\bm{M}$, called the Chern number
of the $n$-th state in the GS manifold. The Chern number
$C^{\,}_{n}(\bm{M})$
is a well-defined topological invariant of the bundle
$\{\ket{\Psi^{\,}_{n}(\bm{\phi},\bm{M})}\}$ of
many-body eigenstates over the fBZ, provided that the energy level that
corresponds to $\ket{\Psi^{\,}_{n}(\bm{\phi},\bm{M})}$
does not cross any other level in
the fBZ~\cite{Thouless1982,Niu1985a,Niu1985a,Thouless1989}.

That the dimensionless quantum Hall conductivity
on the left-hand side of
Eq.\ (\ref{eq: Niu Thouless formula a})
must be a rational number in a FQH phase
in the thermodynanic limit~\cite{Laughlin1981}
is understood from the fact that
the many-body Chern number (\ref{eq: Niu Thouless formula b})
is integer valued. Correspondingly, the right-hand side of
Eq.\ (\ref{eq: Niu Thouless formula a})
is a non-integer rational number if and only if
$N^{\,}_{g=1}>1$ and the sum over all $C^{\,}_{n}$ is not a multiple of
$N^{\,}_{g=1}$.
Moreover, at any quantum phase transition
driven by the parameters $\bm{M}$
by which the Hall conductivity
changes discontinuously,
the many-body Berry curvature
(\ref{eq:berry})
must become singular somewhere in the fBZ.
(This observation in the context of the quantum Hall
and spin quantum Hall effect
was made in Refs.~\onlinecite{Varney2011}
and~\onlinecite{Zeng17}, respectively.
A similar argumentation can also be found
in Refs.~\onlinecite{Carollo2005,Zhu2006}
for topological invariants in quantum spin chains.)
This quantum phase transition is called a plateau transition to emphasize
the fact that the quantum Hall conductivity is constant on either side of
this transition. It is called a topological transition to emphasize
that the phases on either side of this transition differ topologically
through the values taken by the quantum Hall conductivity.

Away from the thermodynamic limit, i.e., when 
$L^{\,}_{1}$ and $L^{\,}_{2}$ are non-vanishing positive integers,
the equality (\ref{eq: Niu Thouless formula a})
is no longer valid.
Indeed, the Hall conductivity, defined through the Kubo formula for
one choice of twisted boundary conditions, need not equal that for
another choice of twisted boundary conditions away from the
thermodynamic limit, contrary to what is implied by
Eq.~(\ref{eq: Niu Thouless formula a})
in the thermodynamic limit.
The GS manifold, if degenerate
in the thermodynamic limit, is generically non-degenerate away from the
thermodynamic limit.
On the other hand, the Chern number
(\ref{eq: Niu Thouless formula b})
remains quantized provided the energy eigenstate
$|\Psi^{\,}_{n}(\bm{\phi},\bm{M})\rangle$
is non-degenerate away from the
thermodynamic limit everywhere in the fBZ and in some region of
parameter space.

In the following,
we shall focus on the case where only one of the $N^{\,}_{g=1}$ GSs
in the thermodynamic limit 
has a non-vanishing Chern number,
\begin{equation}
C^{\,}_{n}(\bm{M})=
\begin{cases}
0,
& \hbox{ if $n\neq n^{\star}=1,\cdots,N^{\,}_{g=1}$,}
\\
C^{\,}_{\star}(\bm{M})\neq0,
& \hbox{otherwise,}
\end{cases}
\label{eq: def index star}
\end{equation}
deep in any insulating phase of parameter space. 
We have verified numerically that all cases studied in this paper
fulfill the condition of Eq.~\eqref{eq: def index star}
that only one state in the GS manifold has a non-vanishing Chern
number, in agreement with previous analytical and numerical
results~\cite{Thouless1989,Sheng2003,Kourtis2012a}.
In cases where more than one states in the GS manifold have a
nonzero Chern number, then Weyl nodes will appear in a sequence of
GS-excited state level crossings. We shall not encounter such a case
in this work.
Now, even though a plateau
transition is rounded by finite-size effects, the Chern number
$C^{\,}_{\star}(\bm{M})$ of the bundle of GSs
$\{|\Psi^{\,}_{n}(\bm{\phi},\bm{M})\rangle\}$
over the fBZ is a discontinuous function
of the parameters $\bm{M}$ that drive the plateau transition in
the thermodynamic limit.
Correspondingly, the Berry curvature
$F^{\,}_{\star}(\bm{\phi},\bm{M})$ computed away from
the thermodynamic limit must develop one or more singularities in the
fBZ for critical values of $\bm{M}$ that drive
the plateau transition in the thermodynamic limit
as illustrated in Fig.\ \ref{fig:MB Berry curvature}.
We shall further \textit{assume} that the jump in
$C^{\,}_{\star}(\bm{M})$ results from a level touching
between two many-body states
$|\Psi^{\,}_{1}(\boldsymbol{M})\rangle$
and
$|\Psi^{\,}_{2}(\boldsymbol{M})\rangle$, i.e.,
from one or more Weyl points (diabolical points) in $\bm{\phi}-\bm{M}$ space
alluded to in Sec.~\ref{sec: Introduction}. In this situation, the Kubo formula~\eqref{eq:berry-alt}
contains a resonant denominator due to the two states participating
in the level crossing. This yields the dominant contribution to the Berry curvature.
This mechanism indeed describes all our exact diagonalization calculations.
What will not be done in this paper is a finite-size scaling analysis
to show that this assumption remains valid as the thermodynamic limit is
approached. Below we reserve the notation $C(\bm{M})\equiv
C^{\,}_{\star}(\bm{M})$ for the single nonzero Chern number in the GS manifold.

\section{FCI-to-trivial transitions}
\label{sec: FCI-to-trivial transitions}

\subsection{Models}

We focus on three widely used models of a single species of
interacting hardcore particles (fermions or bosons) hopping on
two-dimensional lattices made of
$L^{\,}_{1}\times L^{\,}_{2}$ unit cells,
with two sites per unit cell. Their general form is
\begin{subequations}
\begin{equation}
\widehat{H}\:=
\widehat{H}^{\,}_{\textrm{kin}}
+
\widehat{H}^{\,}_{\textrm{int}}.
\end{equation}
The kinetic-energy
$\widehat{H}^{\,}_{\textrm{kin}}$ may be
written in reciprocal space as
\begin{equation}
\widehat{H}^{\,}_{\textrm{kin}}\:=
\sum_{{\bm{k}}\in\textrm{BZ}}
\widehat{\Psi}^{\dag}_{{\bm{k}}}\,\mathcal{H}^{\,}_{\bm{k}}\,\widehat{\Psi}^{\,}_{\bm{k}},
\end{equation} 
where $\widehat{\Psi}^{\dag}_{{\bm{k}}}\equiv
\left(\widehat{c}^{\dag}_{{\bm{k}},A}\,,\widehat{c}^{\dag}_{{\bm{k}},B}\right)$
is the spinor whose two components are made up from two single-particle
creation operators with wave-number $\bm{k}$
in the first Brillouin zone of crystal momenta (not to be confused
with the fBZ of twists in the boundary conditions) and sublattice
index $A$ and $B$, respectively. The single-particle $2\times 2$ matrix
$\mathcal{H}^{\,}_{{\bm{k}}}$ is here defined by
\begin{equation}
\mathcal{H}^{\,}_{{\bm{k}}}\:=
d^{\,}_{0,\bm{k}}\,\tau^{\,}_{0}
+
\bm{d}^{\,}_{\bm{k}}\cdot\bm{\tau}
+
\mu^{\,}_{\mathrm{s}}\,\tau^{\,}_{3},
\end{equation}
\end{subequations}
where $\tau^{\,}_{0}$ and $\bm{\tau}=(\tau^{\,}_{1},\tau^{\,}_{2},\tau^{\,}_{3})$
are the $2\times2$ unit and Pauli matrices in sublattice space, respectively,
and $\mu^{\,}_{\mathrm{s}}$ is a chemical-potential imbalance between
the two sublattices. For both fermions and bosons, the interaction
$\widehat{H}^{\,}_{\textrm{int}}$ is defined to be
\begin{equation}
\widehat{H}^{\,}_{\textrm{int}}\:=
V^{\,}_{1}
\sum_{\langle\bm{i},\bm{j}\rangle}
\widehat{n}^{\,}_{\bm{i}}\,
\widehat{n}^{\,}_{\bm{j}}
+
V^{\,}_{2}
\sum_{\langle\langle\bm{i},\bm{j}\rangle\rangle}
\widehat{n}^{\,}_{\bm{i}}\,
\widehat{n}^{\,}_{\bm{j}},
\end{equation}
where $\widehat{n}^{\,}_{\bm{i}}\:=
\widehat{c}^{\dag}_{\bm{i}}\,\widehat{c}^{\,}_{\bm{i}}$
is the quasi-particle number operator on site $\bm{i}$ of the lattice
and $\langle \dots \rangle$ and $\langle\langle \dots \rangle\rangle$
denote nearest and next-nearest neighbors, respectively.
The sign of the parameters $V^{\,}_{1}\geq0$ and $V^{\,}_{2}\geq 0$ correspond
to repulsive interactions.

\begin{figure}[t]
 \centering
\includegraphics[width=0.99\columnwidth]{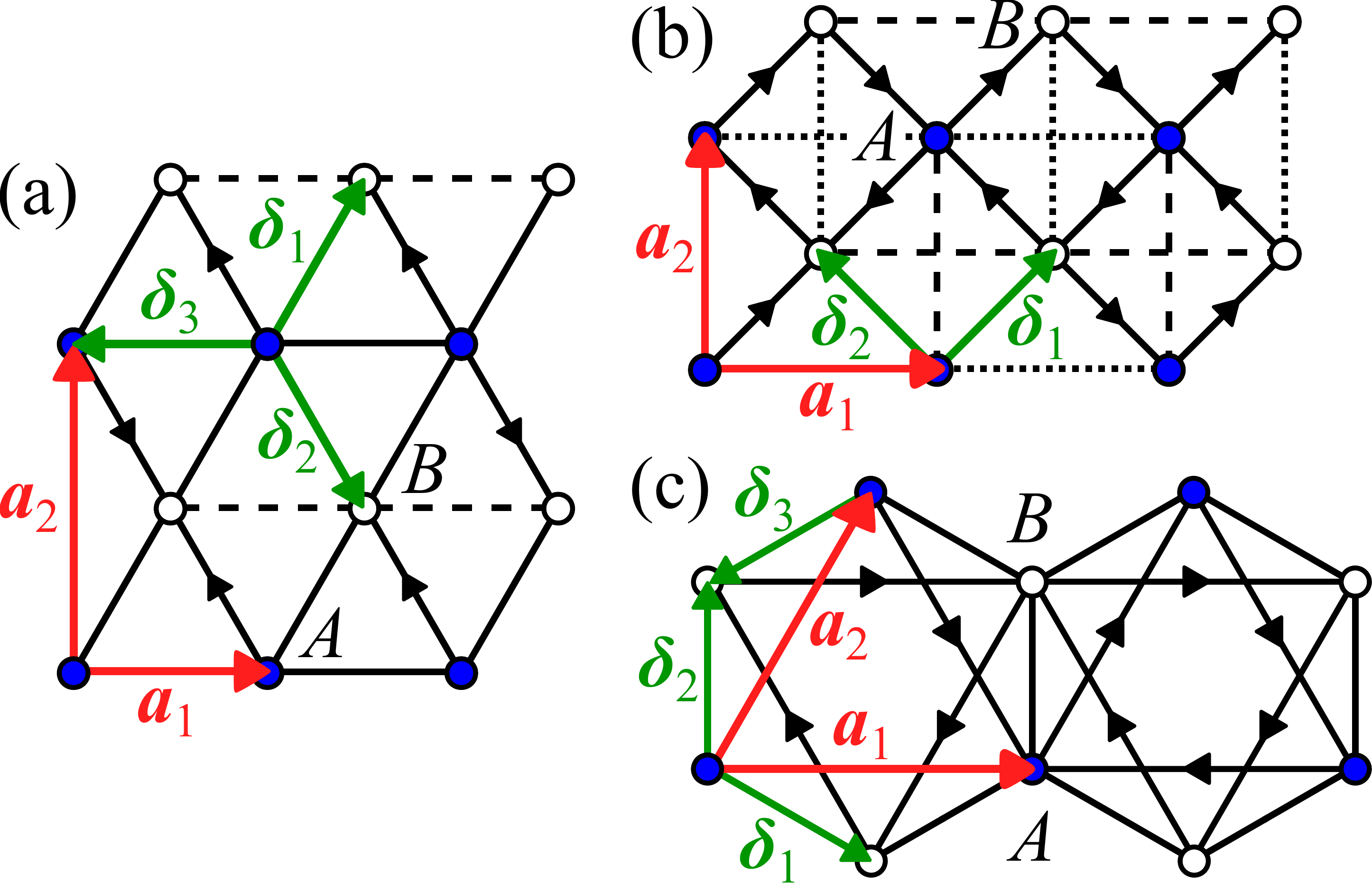}
\caption{
Schematic definition of (a) triangular-lattice model of
Eq.\ (\ref{eq:tri}), (b) checkerboard-lattice model of
Eq.\ (\ref{eq:ckb}), and (c) honeycomb-lattice model of
Eq.\ \eqref{eq:haldane}.
Hoppings in the direction of an arrow add $\varphi$
to the phase of the electron wave function, hopping in the opposite
direction subtracts $\varphi$ ($\varphi=\pi/2$ for triangular-lattice
model, $\varphi=\pi/4$ for checkerboard-lattice model, $\varphi=0.4\pi$ for honeycomb-lattice model); dashed
(dotted) lines denote hoppings with a negative
(positive) sign; in all models third-neighbor hoppings are uniform and
are omitted for clarity.}\label{fig:models}
\end{figure}

The triangular-lattice model of
Refs.~\onlinecite{Venderbos2011a,Kourtis2012a,Kourtis2013a,Kourtis2013}
is defined by choosing
$d^{\,}_{0,\bm{k}}$
and
$\bm{d}^{\,}_{\bm{k}}\equiv
(d^{\,}_{1,\bm{k}},d^{\,}_{2,\bm{k}},d^{\,}_{3,\bm{k}})^{\mathsf{T}}$ as
\begin{subequations}
\label{eq:tri}
\begin{align}
&
d^{\,}_{0,\bm{k}}\:=
2t^{\,}_{3}
\sum_{j=1}^{3}
\cos(2{\bm{k}}\cdot\bm\delta^{\,}_{j}),
\\
&
d^{\,}_{j,\bm{k}}\:=
2t\,\cos({\bm{k}}\cdot\bm\delta^{\,}_{j}),
\qquad j=1,2,3,
\end{align}
\end{subequations}
where
$\bm\delta^{\,}_{1}=(1/2,+\sqrt{3}/2)^{\mathsf{T}}$,
$\bm\delta^{\,}_{2}=(1/2,-\sqrt{3}/2)^{\mathsf{T}}$,
and
$\bm\delta^{\,}_{3}=-
(\bm\delta^{\,}_{1}+\bm\delta^{\,}_{2})$
[see Fig.~\ref{fig:models}(a)]. In all calculations for this
triangular-lattice model, we choose $t>0$ and $V^{\,}_{2}=\mu_{\mathrm{s}}=0$.
The $V^{\,}_{1}$-$t^{\,}_{3}$ phase diagram of this model at density $\rho=1/3$
particles per site has been mapped out in detail in
Refs.~\onlinecite{Kourtis2012a,Kourtis2013} and contains the
competition between a FCI and a charge-density wave (CDW) state.

The model of Refs.~\onlinecite{Neupert2011,Sun2011} is defined on the
checkerboard lattice with primitive vectors $\bm{a}^{\,}_{1} =
(1,0)^{\mathsf{T}}$ and $\bm{a}^{\,}_{2} = (0,1)^{\mathsf{T}}$ as
\begin{subequations}
\begin{align}
&
d^{\,}_{0,\bm{k}}\:=
2t^{\,}_{3}\,
(
\cos 2\bm{k}\cdot\bm\delta^{\,}_{1}
+
\cos 2\bm{k}\cdot\bm\delta^{\,}_{2}
),
\\
&
d^{\,}_{1,\bm{k}}\:=
2t\,
\cos\varphi
(
\cos\bm{k}\cdot\bm\delta^{\,}_{1}
+
\cos\bm{k}\cdot\bm\delta^{\,}_{2}
),
\\
&
d^{\,}_{2,\bm{k}}\:=
2t\,
\sin\varphi
(
\cos\bm{k}\cdot\bm\delta^{\,}_{1}
+
\cos\bm{k}\cdot\bm\delta^{\,}_{2}
),
\\
&
d^{\,}_{3,\bm{k}}\:=
2t^{\,}_{2}\,
(
\cos\bm{k}\cdot\bm{a}^{\,}_{1}
-
\cos\bm{k}\cdot\bm{a}^{\,}_{2}
),
\end{align}
\label{eq:ckb}%
\end{subequations}
where $t$, $t^{\,}_{2}$, and $t^{\,}_{3}$ are nearest-, second
nearest-, and third nearest-neighbor hopping amplitudes, respectively,
and $\bm\delta^{\,}_{1}=(\sqrt{2}/2,\sqrt{2}/2)^{\mathsf{T}}$,
$\bm\delta^{\,}_{2}=(-\sqrt{2}/2,\sqrt{2}/2)^{\mathsf{T}}$
[see Fig.~\ref{fig:models}(b)]. In the following, we fix
$t<0$,
$t^{\,}_{2}/t=-1/(2+\sqrt{2})$,
$t^{\,}_{3}/t=-1/(2+2\sqrt{2})$,
and $\varphi=\pi/4$,
which for $V^{\,}_{1}=V^{\,}_{2}=0$ generates a nearly flat,
topologically nontrivial lower band with a nonzero Chern
number~\cite{Sun2011}. The interacting model at density $\rho=1/6$
particles per site (filling $\nu=1/3$ of the lower band in the
non-interacting limit) has been studied
extensively~\cite{Neupert2011,Sheng2011,Wu2012,Lauchli2012,Kourtis2013a}.
In particular, it was found that (a) the GS of the model
is a topologically ordered FCI for $V^{\,}_{2}=\mu_{\mathrm{s}}=0$ and
$V^{\,}_{1}$ ranging from moderate to
infinite~\cite{Sheng2011,Kourtis2013a}, (b) a large enough
$V^{\,}_{2}$ drives the model into a Fermi liquid-like phase for any
$V^{\,}_{1}>0$~\cite{Sheng2011}, and (c) for large enough
$\mu_{\mathrm{s}}$ the model transitions into a charge-modulated,
topologically trivial state~\cite{Neupert2011,Kourtis2013a}. Our
results are fully consistent with these findings.

Finally, we define the Haldane model on the honeycomb lattice
as~\cite{Haldane1988,Neupert2011}
\begin{subequations}
\label{eq:haldane}
\begin{align}
&
d^{\,}_{0,\bm{k}}\:=
2t^{\,}_{2}\,
\cos\varphi
\sum_{i=1}^{3}\cos\bm{k}\cdot\bm{a}^{\,}_{i},
\\
&
d^{\,}_{1,\bm{k}}\:=
t
\sum_{i=1}^{3}
\cos\bm{k}\cdot\bm{\delta}^{\,}_{i},
\\
&
d^{\,}_{2,\bm{k}}\:=
t
\sum_{i=1}^{3}
\sin\bm{k}\cdot\bm{\delta}^{\,}_{i},
\\
&
d^{\,}_{3,\bm{k}}\:=
-
2t^{\,}_{2}\,
\sin\varphi
\sum_{i=1}^{3}
\sin\bm{k}\cdot\bm{a}^{\,}_{i}
\nonumber\\
&
\hphantom{d^{\,}_{3,\bm{k}}\:=}
+
t^{\,}_{3}
\left[
e^{
\mathrm{i}\bm{k}\cdot(\bm\delta^{\,}_{2}
+
\bm{a}^{\,}_{1})
  }
+
e^{
\mathrm{i}\bm{k}\cdot
(
\bm\delta^{\,}_{2}
-
\bm{a}^{\,}_{1})
}
+
e^{
  -2\mathrm{i}\bm{k}\cdot\bm\delta^{\,}_{2}
}
\right],
\end{align}
\end{subequations}
with $\bm{a}^{\,}_{1}=(\sqrt{3},0)^{\mathsf{T}}$,
$\bm{a}^{\,}_{2}=(\sqrt{3}/2,3/2)^{\mathsf{T}}$,
$\bm{a}^{\,}_{3}=\bm{a}^{\,}_{2}-\bm{a}^{\,}_{1}$,
$\bm\delta_{1} = (\sqrt{3}/2,-1/2)^{\mathsf{T}}$,
$\bm\delta_{2} = (0,1)^{\mathsf{T}}$,
and
$\bm\delta_{3} = (-\sqrt{3}/2,-1/2)^{\mathsf{T}}$
[see Fig.~\ref{fig:models}(c)]. Following Ref.~\onlinecite{Wang2011}, we set
$t^{\,}_{2}/t = 0.60$,
$t^{\,}_{3}/t = 0.58$,
$\mu_s = 0$,
and
$\varphi = 0.4\pi$. The bosonic version of this model has a $\nu=1/2$
FCI GS for any $V^{\,}_{1} \geq 0$, but transitions to a CDW
when $V^{\,}_{2}>V^{\,}_{1}$~\cite{Wang2011}.

The ranges of parameters for all three models
are chosen so as to easily identify and characterize quantum
phase transitions. As long as point-like degeneracies (nodal points)
appear in the fBZ for a critical value of some parameter,
then our methodology is applicable
and small changes in model parameters are inconsequential.

The many-body Berry curvature can be evaluated accurately for finite
clusters using the Lanczos
method~\cite{Venderbos2011a,Kourtis2012a,Kourtis2013a,Kourtis2013}.
In the examples studied in this work, we study the topological
phase transition between a degenerate FCI state, with an accurately
quantized topological invariant $C\not=0$,
and a topologically trivial state with $C=0$.
Depending on geometrical details, the FCI GSs may or
may not all be in the same symmetry sector~\cite{Regnault2011}. In the
former case, there is an energy splitting between quasi-degenerate GSs
of the FCI phase in finite-size numerics and it is expected that only
one of the GSs that we shall label by $n=n^{\,}_{\star}$
will carry the entire topological response~\cite{Thouless1989,Sheng2003}.
Here we focus on this case and
therefore all the results we present are for the symmetry sector that
contains the GSs. In the $\nu=2/3$ FCI phase of the fermionic
triangular-lattice model, one of the three quasi-degenerate/finite-size split FCI states has $C^{\,}_{\star}=2$,
whereas the other two have $C=0$ (average $C$ is 2/3,
as expected for a fractional quantum Hall state at this filling),
while in the $\nu=1/3$ FCI phase of the fermionic
checkerboard-lattice model one GS has $C^{\,}_{\star}=1$
and the other two have
$C=0$ (average $C$ is 1/3). The GS of the bosonic Haldane
model at $\rho=1/4$ is a twice-(quasi-)degenerate $\nu=1/2$ FCI state,
with one of the two states in the GS manifold having $C^{\,}_{\star}=1$
and the other $C=0$ (average $C$ is 1/2).
For all models, we investigate a topological phase
transition out of the FCI phase upon variation of parameters;
$V^{\,}_{1}$ and $t^{\,}_{3}$ in the triangular-lattice model,
$V^{\,}_{2}$ and $\mu_{\mathrm{s}}$ in the checkerboard-lattice model,
and $V^{\,}_{2}$ in the Haldane model. We have evaluated $C$ on both
sides of the transitions by performing the integration in
Eq.~\eqref{eq: Niu Thouless formula b} numerically on finite grids of
the fBZ of sizes up to $64\times64$ and have obtained accurately
quantized values, in agreement with the expected behavior.

\begin{figure*}[t]
 \centering
\includegraphics[width=0.99\textwidth]{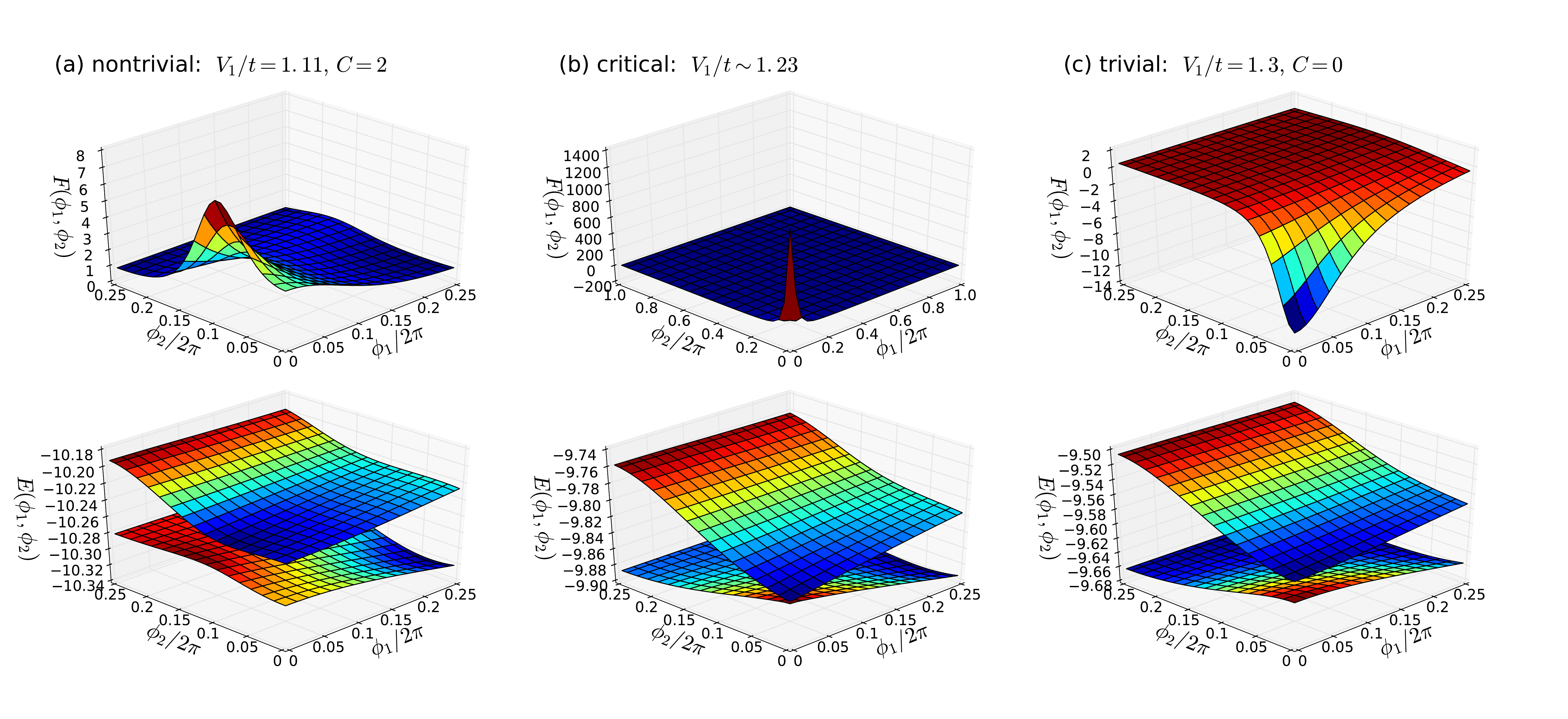}
\caption{
Many-body Berry curvatures (top panels)
and many-body energy dispersions in the fBZ
(bottom panels) for the FCI and CDW states of the
fermionic model defined by Eqs.~\eqref{eq:tri}
on a triangular lattice made of 18 sites
at the fermionic density
(number of fermions divided by the number of sites)
$\rho=1/3$
holding $t^{\,}_{3}/t=0.22$ fixed.
In panel (a),
$V^{\,}_{1}/t=1.11$ (topologically nontrivial $\nu=2/3$ FCI phase).
In panel (b),
$V^{\,}_{1}/t=1.23$ (close to the transition point).
In panel (c)
$V^{\,}_{1}/t=1.3$ (topologically trivial CDW phase).
        }
\label{fig:gap_berry_tri}
\end{figure*}

\subsection{Nodes between many-body energy levels}
\label{sec:crossings}

The generalization of nodal points found in
non-interacting band structures to interacting systems occurs when
the many-body gap closes at isolated points in the fBZ.
As we detail in Appendix~\ref{sec:Symm}, spatial symmetries of a system
with periodic boundary conditions translate to symmetries in the fBZ
when twisted boundary conditions are imposed.
The effect of symmetries on the energy eigenvalue spectrum
in the fBZ has also been discussed in Ref.\ \onlinecite{Varney2011}.
There, it is argued that the Chern number
(\ref{eq: def index star})
for a many-body interacting Hamiltonian with inversion symmetry
can only change by an odd integer if the locations in the fBZ
at which the level crossing takes place is an inversion-symmetric point.
We shall encounter other symmetries that constrain the points
in the fBZ at which diabolical points must occur. For instance, it was shown in
Ref.~\onlinecite{Fang2012} that Weyl nodes on a point in momentum
space whose little group contains a four-fold or six-fold rotation,
can be of higher charge. Similarly,
if an interacting many-body system has
a three-fold rotation symmetry, band touchings at high-symmetry points
in the fBZ may be of higher charge.  We will discuss one such example
below.

First, we focus on the critical point between $\nu=2/3$ FCI and CDW
phases of the fermionic triangular-lattice model of
Eqs.\ \eqref{eq:tri} which is a singly-charged Weyl node. In this
example, upon fine-tuning in parameter space to criticality, the pair
of many-body energy dispersions in the fBZ for the many-body state
supporting the many-body Chern number $C^{\,}_{\star}=2$, on the one
hand, and that for the many-body state supporting a CDW, on the other
hand, touch at the two inequivalent points
$\bm{\phi}\equiv(\phi^{\,}_{1},\phi^{\,}_{2})=(0,0)$ and
$\bm{\phi}=(0,\pi)$.  Figure \ref{fig:gap_berry_tri} demonstrates this
level touching.  Figure \ref{fig:gap_berry_tri} also displays the
dependence of the energy dispersion of these many-body states on the
twisted boundary conditions.  We note that: (i) the dispersion in the
fBZ of this pair of many-body states is linear around the nodal
points, (ii) the Berry curvatures of this pair of many-body states
have the same sign at both points in the fBZ, and (iii) the presence
of two points in the fBZ at which the Berry-curvature diverges leads
to a jump of magnitude 2 in the Chern numbers of the two states
involved in the transition. The transition in the fermionic
checkerboard-lattice model of Eqs.~\eqref{eq:ckb} has the same
features, with the only difference that for this model there is only
one non-equivalent nodal point, namely at the center $\bm{\phi}=(0,0)$
of the fBZ, and the corresponding jump in the magnitude of the Chern
number of the GS manifold upon crossing the quantum critical
point in parameter space is thus 1.

\begin{figure}[t]
\centering
\includegraphics[width=0.99\columnwidth]{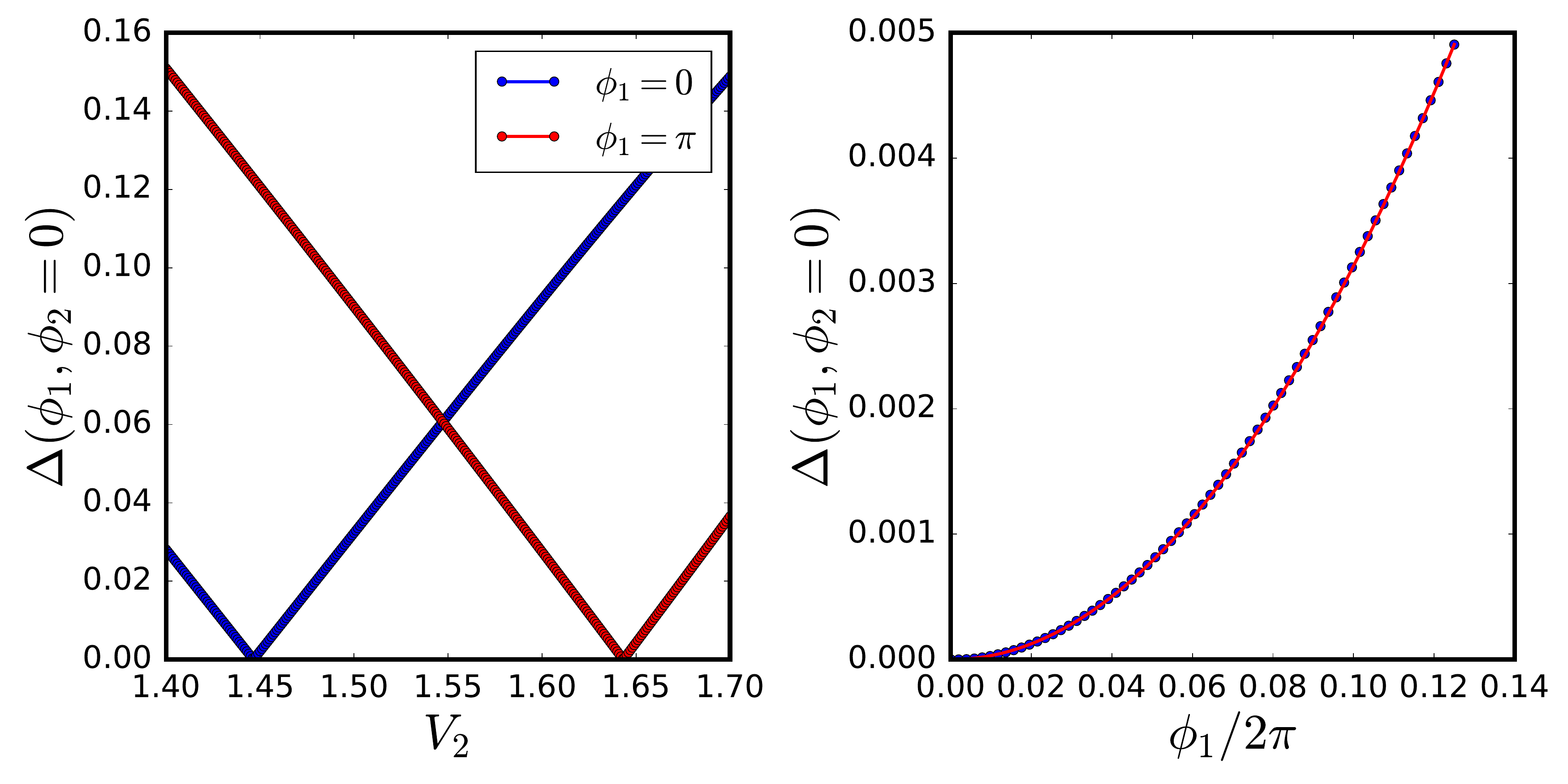}
\caption{
Many-body energy gaps between the FCI
and CDW energy levels
for the bosonic interacting Haldane model of Ref.~\onlinecite{Wang2011}
on a honeycomb lattice made of 32 sites,
as defined in Eq.~(\ref{eq:haldane}).
The average density of hardcore bosons per site is $\rho=1/4$ and $V^{\,}_{1}=0$.
In panel (a), $V^{\,}_{2}$ is varied holding
$\bm{\phi}\equiv(\phi^{\,}_{1},\phi^{\,}_{2})$ fixed to either
$\bm{\phi}=(0,0)$ or $\bm{\phi}=(\pi,0)$.
Panel (a) shows that the many-body gap closes at
$V^{\,}_{2}\sim 1.446$ and $1.642$, respectively.
In panel (b), $\phi^{\,}_{1}$ is varied holding
$V^{\,}_{2}=1.446$ and $\phi^{\,}_{2}=0$ fixed.
The red line in (b) is the fit
$\beta_{1}\,\phi^{2}_{1}$ with $\beta_{1}\simeq\pi/10$.
}
\label{fig:gap_honey}
\end{figure}

\begin{figure}[t]
\centering
\includegraphics[width=0.99\columnwidth]{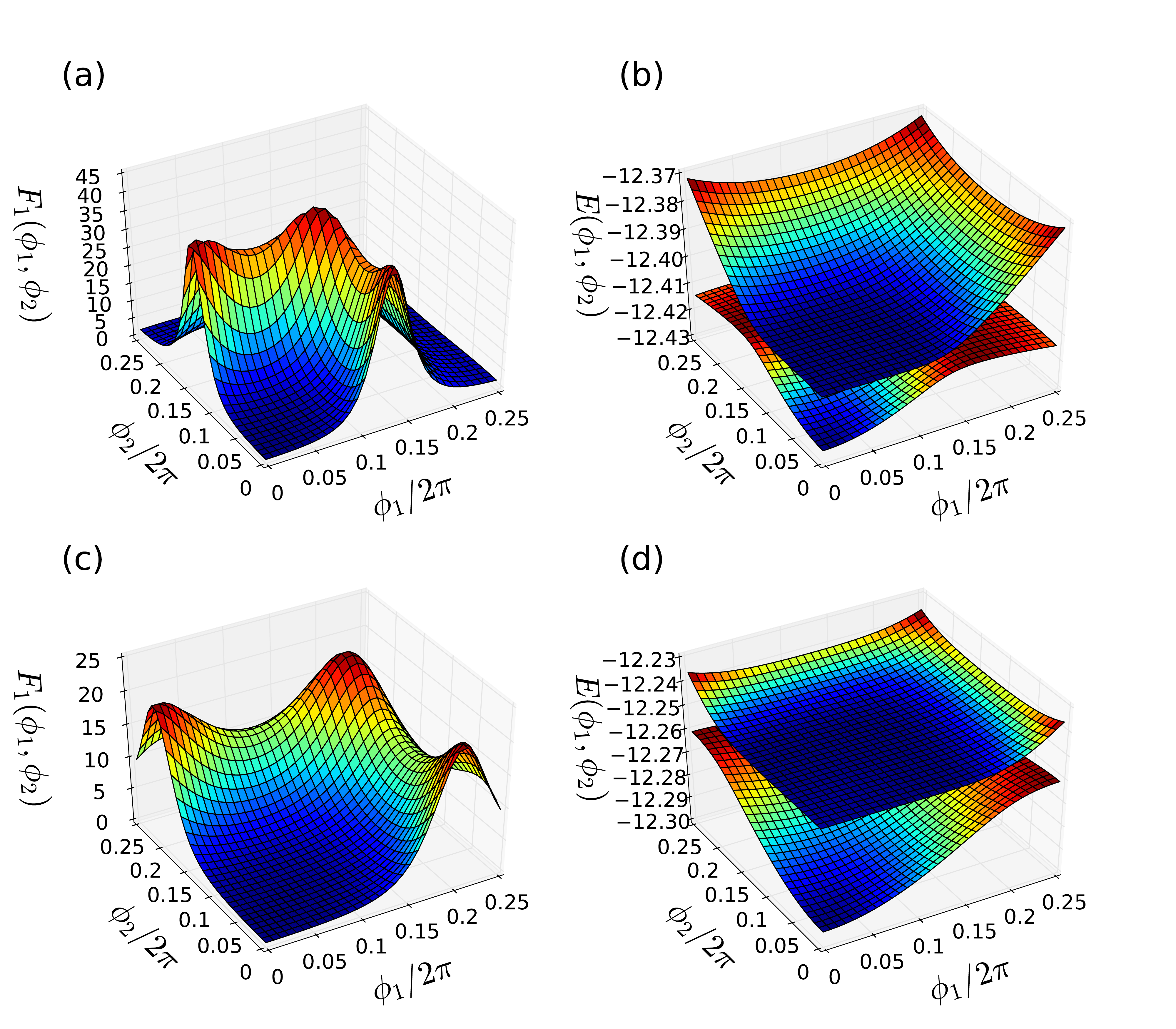}
\caption{
(a,c) Berry curvature of the state $\ket{\Psi^{\,}_{1}}$ corresponding to
first-excited energy level $E^{\,}_{1}$ and (b,d) energy levels $E^{\,}_{1}$
and $E^{\,}_{2}$ in flux space for a 32-site cluster of the interacting
Haldane model of Ref.~\onlinecite{Wang2011}, as defined in
Eq.~\ref{eq:haldane}, at density $\rho=1/4$ hardcore bosons per site
with $V^{\,}_{1} = 0$ and (a,b) $V^{\,}_{2}/t = 1.48$, which is close to the critical point, and (c,d)
$V^{\,}_{2}/t = 1.52$ that is far away from it.
        }
\label{fig:gap_berry_honey}
\end{figure}

The bosonic interacting Haldane model at density $\rho=1/4$
(number of particles divided by the number of sites)
exhibits an instance of higher-order touching of many-body energies
in the fBZ when crossing in parameter space
through the transition from
a $\nu=1/2$ bosonic FCI GS manifold to a charge-ordered
GS manifold upon increasing second-neighbor repulsion
$V^{\,}_{2}$~\cite{Wang2011}.
The phenomenology at
this transition as a function of increasing
$V^{\,}_{2}$ in parameter space is the following.
(1) For small $V^{\,}_{2} \lesssim 1.446$
the FCI and CDW many-body energy levels are fully separated in energy,
with the $C^{\,}_{\star}=1$
FCI many-body state being higher in energy than its
quasi-degenerate counterpart (which has $C=0$)
and lower in energy than the CDW many-body state.
The many-body Berry curvatures are smooth functions everywhere
in the fBZ.
(2) When $V^{\,}_{2}$ reaches the lower critical value $V^{\,}_{2} \sim 1.446$,
the many-body $C^{\,}_{\star}=1$ FCI and CDW dispersions in the fBZ
touch at $\bm{\phi}\equiv(\phi^{\,}_{1},\phi^{\,}_{2})=(0,0)$,
\textit{without} a concomitant divergence of the
many-body Berry curvature at that point,
and disperse quadratically with the deviations 
$\phi^{\,}_{1}$ and $\phi^{\,}_{2}$
away from $\phi^{\,}_{1}=0$ and $\phi^{\,}_{2}=0$
as shown in Fig.~\ref{fig:gap_honey}.
(3) For $1.446 \lesssim V^{\,}_{2}\lesssim 1.642$
the two many-body levels near-avoid one another
-- evidently due to many-body level repulsion --
at three points in the fBZ that move away from
$\bm{\phi}=(0,0)$ along high-symmetry lines upon
increasing $V^{\,}_{2}$, while the many-body Berry curvature develops three
corresponding maxima at the same points, see
Fig.~\ref{fig:gap_berry_honey}.
(4) When $V^{\,}_{2}$ reaches the
upper critical value $V^{\,}_{2}\sim 1.642$, the previously-avoided
many-body crossings reach the points
$\bm{\phi}=(\pi,0)$,
$\bm{\phi}=(0,\pi)$, and
$\bm{\phi}=(\pi,\pi)$,
at which the many-body gap now vanishes exactly
and the many-body Berry curvature diverges.
(5) For $V^{\,}_{2}\gtrsim 1.642$ the many-body spectrum
becomes gapped again, with the many-body
$C^{\,}_{\star}=1$ FCI and $C=0$
CDW levels having exchanged places along the many-body energy axis.

Although the phase encountered for
$1.446\lesssim V^{\,}_{2}\lesssim1.642$
is characterized by avoided many-body level crossings,
the Chern numbers of the corresponding many-body states
are nonetheless well-defined,
since the energy spectrum is actually gapped, albeit by finite-size
effects, see Fig.~\ref{fig:gap_berry_honey}. For $V^{\,}_{2}<1.446$,
we find the many-body Chern numbers of the FCI and CDW
states to be 1 and 0, respectively.
For $1.446 \lesssim V^{\,}_{2} \lesssim 1.642$ the many-body
Chern numbers of the ``hybridized'' states, whose levels repel one
another and are separated by the finite-size gap, jump to the values 3
and $-2$, respectively.
 This behavior has been previously observed
for avoided crossings in a disordered FCI model~\cite{Kourtis2012a}
and indicates that the Chern number is mathematically well-defined in finite
clusters with finite-size gaps, even though the energy spectrum is
expected to become gapless in the thermodynamic limit. Note that the
first transition at $V^{\,}_{2}\sim 1.446$ is accompanied by
quadratically dispersing many-body energy levels
around a single nodal point from the fBZ and
a jump of magnitude 2 in the many-body Chern numbers,
whereas the second transition at $V^{\,}_{2} \sim 1.642$ is accompanied by
linearly dispersing many-body energy levels around
three nodal points from the fBZ and a jump of magnitude 3 in
many-body Chern numbers.

\subsection{Phenomenological classification by dispersion in the fBZ}

The cases detailed above suggest analogies between the interacting
systems studied here and Weyl semimetals.

In the above examples, the many-body Berry curvature of a state is fully
parametrized by a vector $\bm{\phi}$ of boundary twists and a vector $\bm{M}$
of parameters as
$F^{\,}_{n}(\bm{\phi},\bm{M})$.
This many-body Berry curvature integrates to a quantized Chern
number in the $\bm{\phi}$-plane for
$\bm{M}\not=\bm{M}^{\,}_{\mathrm{c}}$.
Finally, there is a jump in the many-body
Chern number that depends on the number and
order of level touchings that occur at $\bm{M}=\bm{M}^{\,}_{\mathrm{c}}$.

In Weyl semimetals, the single-particle Berry curvature is fully parametrized
by a vector of ``in-plane'' momenta $\bm{k}^{\,}_{\parallel}$
and an ``out-of-plane'' momentum $k^{\,}_{\perp}$
as $F^{\,}_{n}(\bm{k}^{\,}_{\parallel},k^{\,}_{\perp})$.
This single-particle Berry curvature
integrates to a quantized Chern number in the $\bm{k}^{\,}_{\parallel}$-plane
for $k^{\,}_{\perp}\not=k^{\,}_{\perp,\mathrm{c}}$. Finally, there is a jump in the single-particle Chern number that depends on the number and order of level touchings (Weyl nodes) that occur across $k^{\,}_{\perp}\not=k^{\,}_{\perp,\mathrm{c}}$.

It is known that higher-order Weyl nodes are only possible in the presence of
additional point-group symmetries and split into single Weyl nodes
when the symmetry protecting them is broken~\cite{Wang2011,Fang2012}.
In an analogous fashion, we observe that the quadratic nodal point
appearing at $\bm{\phi}=(0,0)$ in the interacting Haldane
model splits into two linear nodes upon lowering
the rotational symmetry
by introducing a small mass imbalance between sublattices $A$ and $B$.

In non-interacting systems, the presence of nodes in a band structure
at isolated points in $\bm{k}$-space can be a priori determined via
symmetry arguments~\cite{Bradlyn2017}. Whether the many-body energy
landscape in $\bm{\phi}$-space of an arbitrary interacting
system contains nodal points or not is less straightforward to
determine. Even though, in this work, our approach to answering this
question has been trial-and-error, some arguments hold generally. 
Gaplessness can appear at high-symmetry points (HSPs) in the fBZ
associated with space-group symmetries, or in the form of accidental band crossings at arbitrary points in the fBZ~\cite{Thouless1989,Oshikawa2000,Zhang2010a}. However, any degeneracies away from HSPs need to
also obey all space-group symmetries, which means they must come in
multiples. For example, if a degeneracy occurs at an arbitrary
(non-high symmetry) point in the fBZ of a system with $C^{\,}_{6}$ symmetry, for instance,
then there necessarily need to be 5 more copies of this
degeneracy~\cite{Varney2011}. Furthermore, symmetry considerations do not preclude an
exact degeneracy of many-body levels leading to a divergence of the
Berry curvature on a one-dimensional manifold in the fBZ, with a
$\bm{\phi}$-space dispersion resemblant of non-interacting
nodal-line semimetals. We shall not address this case here.

For all the three models studied in Sec.\ \ref{sec:crossings}, near
the critical point $\bm{M}^{\,}_{\mathrm{c}}$ and close to the nodal
point $\bm{\phi}^{\,}_{\mathrm{c}}$, the many-body gap between the
many-body energy levels
$E^{\,}_{1}(\bm{\phi},\bm{M})$ and $E^{\,}_{2}(\bm{\phi},\bm{M})$
corresponding to topologically nontrivial and trivial
many-body states can be
heuristically fitted by the simple form
\begin{align}
&
\left(
E^{\,}_{2}-E^{\,}_{1}
\right)(\bm{\phi},\bm{M})\approx
\nonumber\\
&\qquad\qquad
\sqrt{
\beta^{\,}_{1}(\delta\phi^{\,}_{1})^{2p}
+
\beta^{\,}_{2}(\delta\phi^{\,}_{2})^{2p}
+
\beta^{\,}_{3}|\bm{M}-\bm{M}_{\rm c}^{\star}|^{2}
     },
\label{eq:fitting_dispersion}
\end{align}
where $\bm{M}^{\star}_{\mathrm{c}}$
is the point on the $(m-1)$-dimensional phase boundary that is closest to
$\bm{M}$ and
where $\delta\phi^{\,}_{1}$ and $\delta\phi^{\,}_{2}$
are sufficiently small coordinates measured relative to
$\bm{\phi}^{\,}_{\mathrm{c}}\equiv
(\phi^{\,}_{\mathrm{c}1},\phi^{\,}_{\mathrm{c}2})^{\mathsf{T}}$,
while
$\beta^{\,}_{1}$, $\beta^{\,}_{2}$, and $\beta^{\,}_{3}$
are fitting parameters,
and $\bm{M}$ is sufficiently close to $\bm{M}^{\star}_{\mathrm{c}}$.
For all the nodal points we have encountered in our calculations, the power $p_d$
is always found to be an \textit{integer}, and it is equal to 1
for divergences at HSPs in the triangular and the checkerboard models,
and 2 (1) for the critical point at $V_{2}\sim 1.446$ ($V_{2}\sim
1.642$) in the honeycomb model. By monitoring numerically the Chern
number $C$ of the state corresponding to the lower of the two levels
as the system undergoes the level crossing upon variation of $\bm{M}$,
we find that in all these cases the jump $\Delta C$
in the Chern number obeys the
empirical relation
\begin{equation}
\Delta C= 
\sum_{d=1}^{N^{\,}_{\mathrm{div}}}
p^{\,}_{d}\,,
\label{eq:jump_Chern_number}
\end{equation}
where $N^{\,}_{\mathrm{div}}$ denotes the number of nodes that
develop at points $\bm\phi^{\,}_{\mathrm{c},d}$
with $d=1,\cdots,N^{\,}_{\mathrm{div}}$ in the fBZ as one
approaches a critical point, and $p^{\,}_{d}$ is the corresponding
integer governing the dispersion in $\bm{\phi}$ close to the $d$-th
HSP. Interestingly,
this empirical result mirrors what happens in non-interacting systems,
where Eqs.~\eqref{eq:fitting_dispersion}
and~\eqref{eq:jump_Chern_number} can be rigorously related to one
another.~\cite{Wan2011,Wang2011,Fang2012,Chen2017}
In all the transitions we have investigated, the Berry curvature
always has the \textit{same} sign at all the HSPs of the fBZ
where it becomes divergent simultaneously.

\section{Berry curvature renormalization group}
\label{sec:BCRG}

The similarities that we have observed
in Sec.\ \ref{sec: FCI-to-trivial transitions}
when comparing the many-body Berry curvatures to
the single-particle Berry curvatures
of non-interacting semi-metals close to Weyl nodes
motivate us to develop a scaling
procedure for the many-body Berry curvature that is the many-body
counterpart to the one applied to the single-particle
Berry curvature in
Refs.\ \onlinecite{Chen2016,Chen2016b,Chen2017}.
We then apply this scaling procedure to the
FCI-to-trivial transitions presented in
Sec.\ \ref{sec: FCI-to-trivial transitions}.
We detail the protocol below,
which we refer to as Berry curvature renormalization group
(BCRG), for the two scenarios $p_d=1$ and $p_d=2$ uncovered in
Sec.\ \ref{sec: FCI-to-trivial transitions}.

The procedure we introduce relies on positing
the functional form for the singularity
of the Berry curvature $F^{\,}_{\star}(\bm{\phi},\bm{M})$
and the applicability of a scaling relation that ties
the dependence of $F^{\,}_{\star}(\bm{\phi},\bm{M})$
on $\bm{\phi}$ to that on the parameters $\bm{M}$
that drive the plateau transition in the thermodynamic limit.
[The index $\star$ in $F^{\,}_{\star}(\bm{\phi},\bm{M})$
was defined below Eq.\ (\ref{eq: def index star}).]
Our approach is phenomenological in that
we do not justify the validity of these assumptions
on the basis of general principles.
Instead, we confirm the fulfillment of these conditions on a case by case basis
with exact diagonalization.

\subsection{Strategy}
\label{sec:Strategy}

The BCRG approach relies on the following three assumptions.

\textbf{Assumption 1 --}
The many-body Berry curvature is
an even function of $\bm{\phi}$
around all the high symmetry points (HSPs) defined by
$\bm{\phi}^{\,}_{\mathrm{c}}=-\bm{\phi}^{\,}_{\mathrm{c}}$
modulo $(2\pi\,n^{\,}_{1},2\pi\,n^{\,}_{2})$
for any pair of integers $n^{\,}_{1}$ and $n^{\,}_{2}$, i.e.,
\begin{equation}
F^{\,}_{\star}(
\bm{\phi}^{\,}_{\mathrm{c}}+\delta\bm{\phi},\bm{M}
         )=
F^{\,}_{\star}(
\bm{\phi}^{\,}_{\mathrm{c}}-\delta\bm{\phi},\bm{M}
         ),
\label{eq: Assumption 1}
\end{equation}
where the Berry curvature $F^{\,}_{\star}(\bm{\phi}^{\,}_{\mathrm{c}},\bm{M})$
was defined implicitly by Eq.\ (\ref{eq: def index star}).

If there exists a quantum critical point
$\bm{M}^{\,}_{\mathrm{c}}$
across which the quantum Hall conductivity 
(\ref{eq: Niu Thouless formula a})
changes discontinuously, we then
posit either one of the following two scenarios upon approaching
in parameter space the quantum critical point
$\bm{M}^{\,}_{\mathrm{c}}$.

\textbf{Assumption 2.1 --} The Berry curvature
(\ref{eq:berry})
displays a local extremum at $\bm{\phi}^{\,}_{\mathrm{c}}$
as a function of $\bm{\phi}$, holding
$\bm{M}$ fixed and close to $\bm{M}^{\,}_{\mathrm{c}}$.
This extremum changes from being a local maximum to a local minimum
as $\bm{M}$ is varied across $\bm{M}^{\,}_{\mathrm{c}}$.
The Berry curvature
$F^{\,}_{\star}(\bm{\phi}^{\,}_{\mathrm{c}},\bm{M})$
diverges as $\bm{M}\to\bm{M}^{\,}_{\mathrm{c}}$.
For example, the dependence on
$\delta\bm{\phi}$ given by the Ornstein-Zernike (OZ) scaling function
\begin{equation}
F^{\,}_{\mathrm{OZ}}(\bm{\phi}^{\,}_{\mathrm{c}}+\delta\bm{\phi},\bm{M})\:=
\frac{
F^{\,}_{\star}(\bm{\phi}^{\,}_{\mathrm{c}},\bm{M})
     }
     {
1
+
\xi^{2}_{1}(\bm{M})\,\delta\phi^{2}_{1}
+
\xi^{2}_{2}(\bm{M})\,\delta\phi^{2}_{2}
     },
\label{eq:Ornstein-Zernike}
\end{equation}
for some dimensionless functions 
$\xi^{\,}_{1}(\bm{M})$
and
$\xi^{\,}_{2}(\bm{M})$, satisfies this assumption.
It then follows that both
$\xi^{2}_{1}(\bm{M})$
and
$\xi^{2}_{2}(\bm{M})$
must share the same singular behavior with
the Berry curvature (\ref{eq:berry})
when $\bm{\phi}$ is held fixed at
$\bm{\phi}^{\,}_{\mathrm{c}}$
while $\bm{M}\to\bm{M}^{\,}_{\mathrm{c}}$.
We shall subsequently call this assumption the peak-divergence scenario.

\begin{figure}[t]
\includegraphics[clip=true,width=0.95\columnwidth]{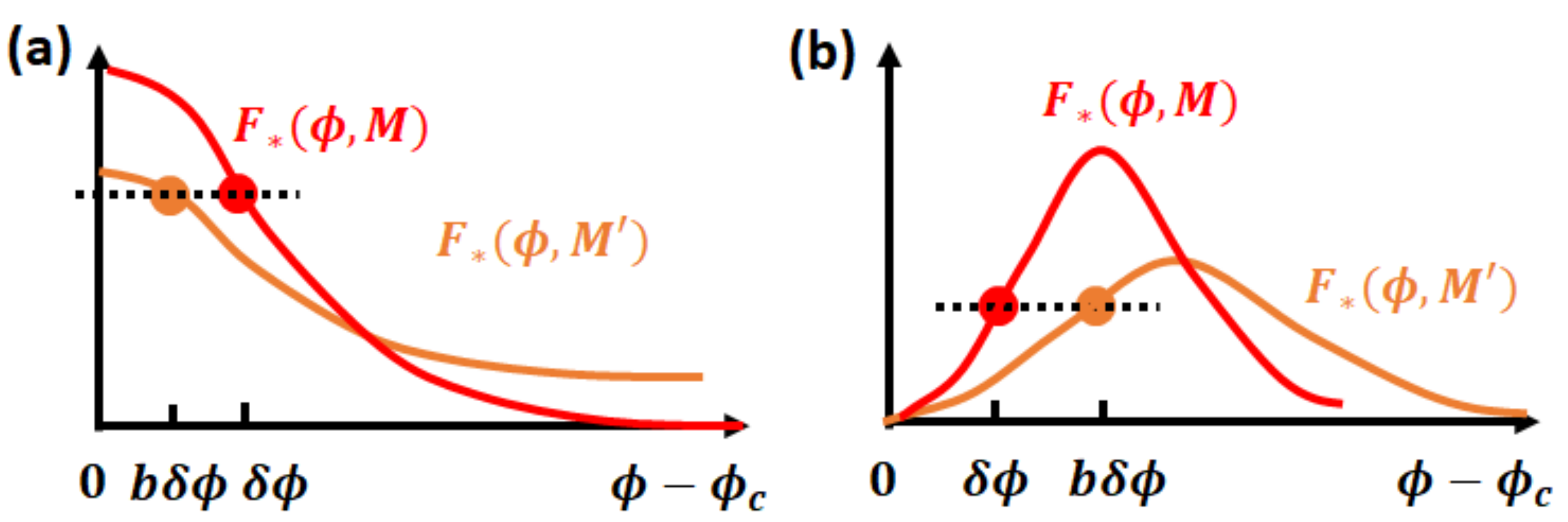}
\caption{
(a) and (b) illustrate
the scaling procedure for Assumption \textbf{2.1} (peak-divergence scenario)
and \textbf{2.2} (ring-divergence scenario), respectively,
which demands the red dot to be equal to the orange dot to solve for
the $\bm{M}^{\prime}$, as indicated by the dashed line. The
profile of Berry curvature evolves from the red lines that have a
large divergence to the orange lines that have a small divergence
under this procedure, without changing the topological invariant,
hence the system flows away from the critical point.
        }
\label{fig:correlation_fn}
\end{figure}

\textbf{Assumption 2.2 --}
The Berry curvature (\ref{eq:berry})
has a continuous mountain ridge (valley)
that surrounds
$\bm{\phi}^{\,}_{\mathrm{c}}$ as a function of
$\bm{\phi}$, holding
$\bm{M}$ fixed and close to $\bm{M}^{\,}_{\mathrm{c}}$.
These extrema change from being local maxima to local minima
as $\bm{M}$ is varied across $\bm{M}^{\,}_{\mathrm{c}}$.
These extrema collapse to
$\bm{\phi}^{\,}_{\mathrm{c}}$
and the Berry curvature at any point along these extrema
diverges as $\bm{M}\to\bm{M}^{\,}_{\mathrm{c}}$. 
We shall subsequently call this assumption the ring-divergence scenario.

\textbf{Assumption 3 --}
For any
$\delta\bm{\phi}\equiv
|\delta\bm{\phi}|\,\widehat{\delta\bm{\phi}}$
nonvanishing yet not too large,
there exists one and only one pair
$\delta\bm{\phi}^{\prime}\equiv
|\delta\bm{\phi}^{\prime}|\,\widehat{\delta\bm{\phi}}$
and
$\bm{M}^{\prime}$
such that
\begin{equation}
F^{\,}_{\star}(\bm{\phi}^{\,}_{\mathrm{c}}+\delta\bm{\phi},
  \bm{M})=
F^{\,}_{\star}(\bm{\phi}^{\,}_{\mathrm{c}}+\delta\bm{\phi}^{\prime},
  \bm{M}^{\prime}).
\label{eq: scaling relation}
\end{equation}

Defining $|\delta\bm{\phi}^{\prime}|/|\delta\bm{\phi}|\equiv b$, 
scenario \textbf{2.1} demands that $0<b<1$, whereby
$F^{\,}_{\star}(\bm{\phi}^{\,}_{\mathrm{c}},\bm{M})$
is a local maximum as a function of $\bm{\phi}$ holding
$\bm{M}$ fixed. 
Scenario \textbf{2.2} demands that $b>1$,
whereby
$F^{\,}_{\star}(\bm{\phi}^{\,}_{\mathrm{c}}+\delta\bm{\phi},
\bm{M})$
is a mountain ridge (valley).
The intuition for Eq.\ (\ref{eq: scaling relation})
is captured by Fig.\ \ref{fig:correlation_fn}.
The scaling ansatz (\ref{eq: scaling relation})
defines a flow in parameter space from
$\bm{M}$ to $\bm{M}^{\prime}$,
along which the divergence of the Berry curvature is reduced,
and hence the system is moving away from the critical point,
as explained in Appendix \ref{sec:dev-red}.
This flow can be encoded into a differential equation as follows.
We define the scaling direction to be
$\widehat{\delta\bm{\phi}}$.
We define the infinitesimal scaling coordinate to be
$\mathrm{d}\ell\equiv|\delta\bm{\phi}|^{2}$.
We define the infinitesimal change in the parameters to be
$\mathrm{d}\bm{M}\equiv
\bm{M}^{\prime}-\bm{M}$.
We do a Taylor expansion of
Eq.\ (\ref{eq: scaling relation})
in powers of $\delta\bm{\phi}$ and $\mathrm{d}\bm{M}$
and use the assumption (2) that
$\bm{\nabla}^{\,}_{\bm{\phi}}F^{\,}_{\star}=0$
at $\bm{\phi}^{\,}_{\mathrm{c}}$ and $\bm{M}$.
If so, there follows the partial differential equation
\begin{equation}
\left.
\bm{\nabla}^{\,}_{\bm{M}}F^{\,}_{\star}
\right|^{\,}_{\bm{\phi}^{\,}_{\mathrm{c}},\bm{M}}
\cdot
\frac{\mathrm{d}\bm{M}}{\mathrm{d}\ell}=
\left.
\left(\frac{1-b^{2}}{2}\right)
\left(
\widehat{\delta\bm{\phi}}
\cdot
\bm{\nabla}^{\,}_{\bm{\phi}}
\right)^{2}
F^{\,}_{\star}
\right|^{\,}_{\bm{\phi}^{\,}_{\mathrm{c}},\bm{M}}.
\label{eq: generic_RG_eq}
\end{equation}
Assuming that
$\left.\bm{\nabla}^{\,}_{\bm{M}}F^{\,}_{\star}
\right|^{\,}_{\bm{\phi}^{\,}_{\mathrm{c}},\bm{M}}$
is non-vanishing and performing this scaling procedure independently
for each tuning parameter $M^{\,}_{i}$ with $i=1,\cdots,m$,
one can rearrange Eq.~\eqref{eq: generic_RG_eq} to obtain a closed
expression for $\mathrm{d}M^{\,}_{i}/\mathrm{d}\ell$.
The row vector
$\left(
\mathrm{d}M^{\,}_{1}/\mathrm{d}\ell,
\cdots,
\mathrm{d}M^{\,}_{m}/\mathrm{d}\ell,
\right)^{\mathsf{T}}\equiv
(\mathrm{d}\bm{M}/\mathrm{d}\ell)^{\mathsf{T}}$
delivers the BCRG flow in the $m$-dimensional parameter space.
The critical points $\bm{M}^{\,}_{\mathrm{c}}$ are identified
numerically from the BCRG flow as the $(m-1)$-dimensional surface on
which the flow directs away from the point at which the flow rate
diverges. {The fixed points $\bm{M}^{\,}_{\mathrm{f}}$
are the points where the flows vanishes. They
can be either stable or unstable depending on the direction
of the BCRG flow.
In short,
\begin{eqnarray}
&&
\hbox{Critical point: }
\left|\frac{\mathrm{d}\bm{M}}{\mathrm{d}\ell}\right|\rightarrow\infty\;,
\hbox{flow directs away,}
\nonumber \\
&&
\hbox{Stable fixed point: }
\left|\frac{\mathrm{d}\bm{M}}{\mathrm{d}\ell}\right|\rightarrow 0\;,
\hbox{flow directs into,}
\nonumber \\
&&
\hbox{Unstable fixed point: }
\left|\frac{\mathrm{d}\bm{M}}{\mathrm{d}\ell}\right|\rightarrow 0\;,
\hbox{flow directs away.}
\nonumber \\
\label{eq:Mc_Mf_definition}
\end{eqnarray} 
To numerically estimate the partial differential equation
(\ref{eq: generic_RG_eq}), we use the triplet of values
$F^{\,}_{\star}(\bm{\phi}_{\mathrm{c}},\bm{M})$,
$F^{\,}_{\star}(\bm{\phi}^{\,}_{\mathrm{c}}
+
\Delta\phi^{\,}_{i}\widehat{\delta\bm{\phi}},\bm{M})$,
and
$F^{\,}_{\star}(\bm{\phi}^{\,}_{\mathrm{c}},\bm{M}+
\Delta M^{\,}_{i}\widehat{\bm{M}}^{\,}_{i})$.

A couple of remarks are in order. First, this approach implicitly
assumes that the bulk gap continuously reduces as $\bm{M}$ approaches
$\bm{M}^{\,}_{\mathrm{c}}$ and vanishes \textit{exactly} at the
critical point. This gap closing must be accompanied by the divergence
of the many-body Berry curvature, as per Eq.~\eqref{eq:berry}. This is
not necessarily the case for an arbitrary topological phase
transition. On the one hand, states whose levels cross with varying
$\bm{M}$ that are in different symmetry sectors do not give rise to
divergent Berry curvature, despite the resonant denominator in the Kubo
formula, due to
cancellation of matrix elements in the numerator. In our studies, this
is circumvented by choosing finite lattices in which the relevant states
happen to fall in the same symmetry sector. The generic situation of a
physical system with finite disorder that breaks all symmetries
naturally falls into this category as well. On the other hand,
topological phase transitions can also occur without a gap closing
whatsoever.~\cite{Amaricci2015} We have not observed this situation in
any of the systems we have studied. The above limitations
notwithstanding, the BCRG approach provides a useful characterization
tool for the detection and classification of topological phase
transitions into and out of states of interacting systems
characterized by a nontrivial quantum Hall response, as will be shown
below in Sec.~\ref{sec:bcrg_application}.

\subsection{Application to FCI-to-trivial transitions}
\label{sec:bcrg_application}

\subsubsection{Peak-divergence scenario}
\label{sec:peak_divergence}

The triangular and checkerboard lattice models realize
the peak-divergence scenario
described in assumption 2.1 of Sec.~\ref{sec:Strategy},
since the Berry curvature peaks
at one or multiple HSPs. Each such peak in the Berry curvature, such
as the one shown in the top panel of Fig.\ \ref{fig:gap_berry_tri}(b),
is well fitted by the OZ ansatz
\eqref{eq:Ornstein-Zernike}. The Berry curvature (\ref{eq:berry})
displays a local extremum at $\bm{\phi}^{\,}_{\mathrm{c}}$ as a
function of $\bm{\phi}$, holding $\bm{M}$ fixed and close to
$\bm{M}^{\,}_{\mathrm{c}}$.  This extremum changes from being a local
maximum to a local minimum as $\bm{M}$ is varied across
$\bm{M}^{\,}_{\mathrm{c}}$, and the Berry curvature
$F^{\,}_{\star}(\bm{\phi}^{\,}_{\mathrm{c}},\bm{M})$ diverges as
$\bm{M}\to\bm{M}^{\,}_{\mathrm{c}}$.

\begin{figure}
\includegraphics[clip=true,width=0.99\columnwidth]{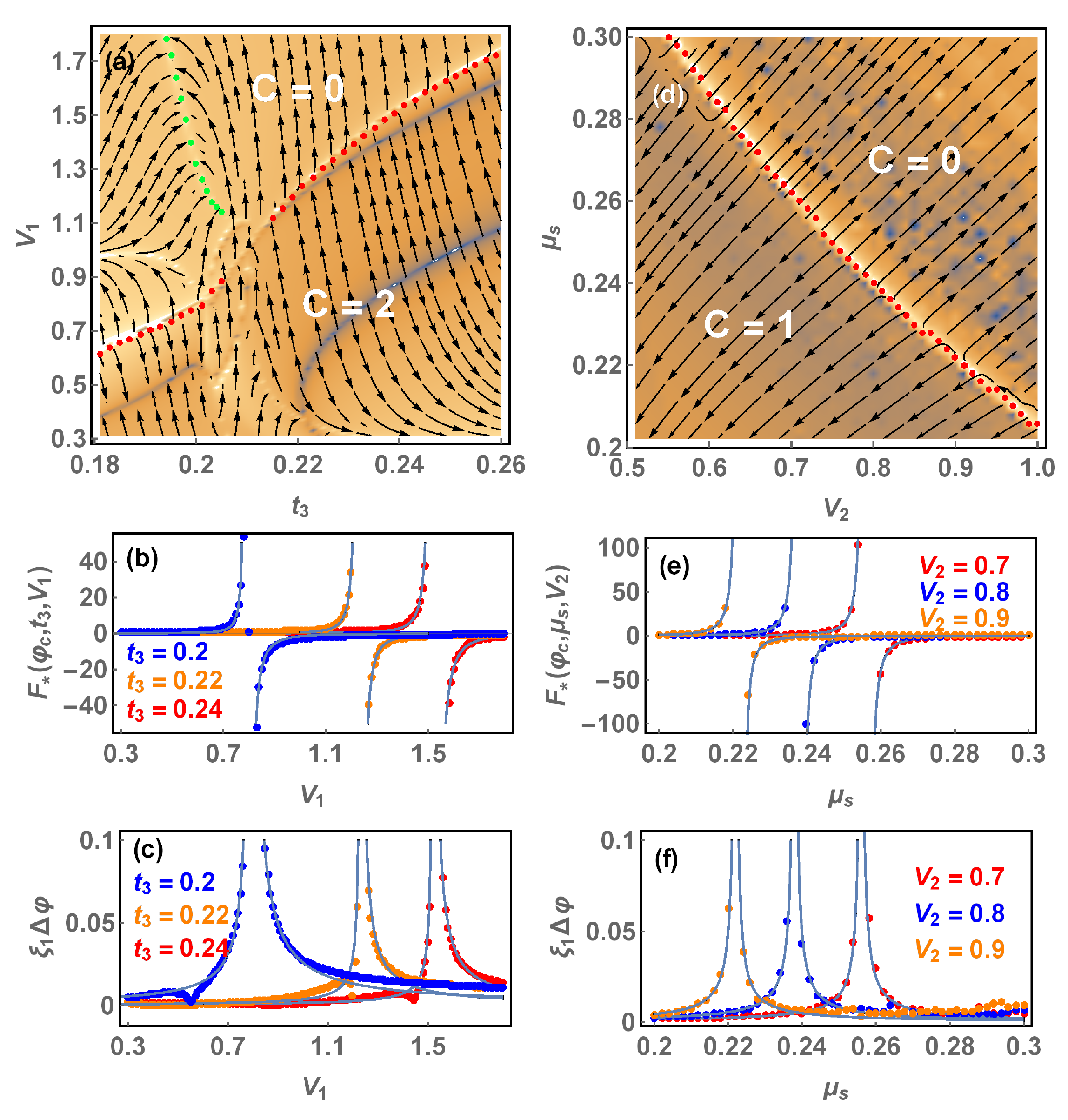}
\caption{
(color online)
(a) BCRG flow for the triangular lattice model using
$\Delta\bm{M}=(\Delta t^{\,}_{3},\Delta V^{\,}_{1})=(0.001,0.01)$,
and, without loss of generality, choosing $b=0$ in
Eq.\ (\ref{eq: scaling relation}),
with the topological invariant $C$ labeled for each phase. 
The flow rate in logarithmic scale is shown by the color code.
Blue color indicates a low flow rate and orange a high flow rate.
Red dots label the phase boundary, and the green dots the line at which
$\bm{\nabla}^{\,}_{\bm{M}}\,F^{\,}_{\star}|^{\,}_{\bm{\phi}^{\,}_{\mathrm{c}},\bm{M}}$
in Eq.\ (\ref{eq: generic_RG_eq}) vanishes.
The scaling ansatz
(\ref{eq:Ornstein-Zernike})
is made to fit the Berry curvature in the neighborhood
of the HSP and of a plateau transition.
(b) and (c) show the divergence of
$F^{\,}_{\star}(\bm{\phi}^{\,}_{\mathrm{c}},t^{\,}_{3},V^{\,}_{1})$
and
$\xi^{\,}_{1}$ versus $V^{\,}_{1}$ at a few selected $t^{\,}_{3}$.
(d) BCRG flow of the checkerboard model using
$\Delta\bm{M}=(\Delta V^{\,}_{2},\Delta\mu^{\,}_{2})=(0.01,0.002)$.
(e) and (f) show the divergence of
$F^{\,}_{\star}(\bm{\phi}^{\,}_{\mathrm{c}},V^{\,}_{2},\mu^{\,}_{2})$
and $\xi^{\,}_{1}$ versus $\mu^{\,}_{2}$ at a few selected $V^{\,}_2$.
The small displacement $\Delta\bm{\phi}=(2\pi/1000,0)$ is used for the calculations.
Both models feature the same
HSP $\bm{\phi}^{\,}_{\mathrm{c}}=(0,0)$, and the divergence
of $F^{\,}_{\star}(\bm{\phi}^{\,}_{\mathrm{c}},\bm{M})$
and $\xi^{\,}_{1}$ are well fitted by the same exponents
$\alpha=2$ and $\nu^{\,}_{1}=1$, as indicated by the solid lines.
}
\label{fig:FCI_tp_V1}
\end{figure}

Numerical results of BCRG applied to the triangular and the
checkerboard lattice model are shown in Fig.\ \ref{fig:FCI_tp_V1}. The
Berry curvature (\ref{eq:berry}) at the HSPs of the fBZ is calculated
by exact diagonalization on a grid over the parameter space of
$\bm{M}$. For the fermionic triangular lattice model at the density
$\rho=1/3$ particles per site, we choose the parameter space
$\bm{M}=(t^{\,}_{3},V^{\,}_{1})$ where $t^{\,}_{3}$ is the
third-neighbor hopping and $V^{\,}_{1}$ is the nearest-neighbor
repulsion.  For the fermionic checkerboard lattice model at the
density $\rho=1/6$, the parameter space is
$\bm{M}=(V^{\,}_{2},\mu^{\,}_{2})$ where $V^{\,}_{2}$ is the
second-neighbor repulsion and $\mu^{\,}_{2}$ is a chemical potential
imbalance between sublattices $A$ and $B$. The numerical result shows
a BCRG flow that correctly captures the phase boundary between FCI and
the topologically trivial state, as indicated by the red dots in Fig.\
\ref{fig:FCI_tp_V1} (a) and (d). The triangular lattice model shows
a rich phase diagram, manifesting both stable and unstable fixed points
[blues lines in Fig.\ \ref{fig:FCI_tp_V1} (a)],
as well as a line (green dots) at which
${\bm\nabla}^{\,}_{\bm{M}}\,F^{\,}_{\star}|^{\,}_{{\bm \phi}_{\mathrm{c}},\bm{M}}$
vanishes and hence higher order expansion of Eq.\
(\ref{eq: generic_RG_eq}) is required to correctly describe the BCRG
equation. Near the fixed point on the phase boundary the BCRG flow becomes chaotic,
and it becomes difficult to extract the phase boundary.

Drawing an analogy to non-interacting
systems~\cite{Chen2017}
and to thermodynamic phase transitions (although we are dealing
with given finite lattices), we denote the ``critical exponents''
associated to the divergences in the many-body Berry curvature
$F^{\,}_{\star}$ and the functions $\xi^{\,}_{1}$ and $\xi^{\,}_{2}$
entering the OZ scaling function
\eqref{eq:Ornstein-Zernike}
by
\begin{subequations}
\label{eq: scaling ansatz Ornstein-Zernike}
\begin{align}
&
F^{\,}_{\star}(\bm{\phi}^{\,}_{\mathrm{c}},\bm{M})\propto
|\bm{M}-\bm{M}^{\,}_{\mathrm{c}}|^{-\alpha},
\label{eq: scaling ansatz Ornstein-Zernike a}
\\
&
\xi^{\,}_{1}(\bm{M})\propto
|\bm{M}-\bm{M}^{\,}_{\mathrm{c}}|^{-\nu^{\,}_{1}},
\qquad
\xi^{\,}_{2}(\bm{M})\propto
|\bm{M}-\bm{M}^{\,}_{\mathrm{c}}|^{-\nu^{\,}_{2}}.
\label{eq: scaling ansatz Ornstein-Zernike b}
\end{align}
\end{subequations}
For these two models, the numerical results are well fitted by
$\alpha\approx 2$ and $\nu^{\,}_{1}\approx\nu^{\,}_{2}\approx 1$. This
behavior mirrors that of Weyl nodes and satisfies the scaling law
$\alpha=\nu^{\,}_{1}+\nu^{\,}_{2}$ introduced in
Appendix \ref{appen:scaling_law}. 
In Appendix \ref{sec:Wannier_correlation_fn},
we provide a complementary interpretation of $\xi^{\,}_{i}(\bm{M})$,
with $i=1,2$, in terms of correlation functions on the lattice
dual to the fBZ.

\subsubsection{Ring-divergence scenario\label{sec:ring_divergence}}

The honeycomb model near $V^{\,}_{2\mathrm{c}}\approx 1.446$, characterized by
quadratic $\boldsymbol\phi$-dispersion around $\boldsymbol\phi_c = (0,0)$,
realizes the ring-divergence scenario, since the
extremum of Berry curvature forms a ring surrounding
$\bm{\phi}^{\,}_{\mathrm{c}}$. The ring is not necessarily circular,
its precise shape is parameter-dependent, and the extremum is
not uniform along the ring, as can be seen in Fig.~\ref{fig:gap_berry_honey}. 

The ``critical'' behavior of the Berry curvature in this scenario is that,
as $\bm{M}\rightarrow\bm{M}^{\,}_{\mathrm{c}}$,
(i) the extremum of the Berry curvature diverges,
(ii) the radius of the ring along which the Berry curvature reaches
its extremal value vanishes,
(iii)
and $F^{\,}_{\star}(\bm{\phi}^{\,}_{\mathrm{c}},\bm{M})$ remains
finite. The extremum changes from a maximum to a minimum as $\bm{M}$
passes $\bm{M}^{\,}_{\mathrm{c}}$. Based on these features, we propose
the same scaling procedure, Eq.~(\ref{eq: scaling relation}), but with
the choice $b>1$ to obtain the BCRG flow, as shown schematically in
Fig.~\ref{fig:correlation_fn}~(b).
The scaling procedure gradually reduces the magnitude of the extremum and
increases the radius of the ring, and hence the system is gradually
flowing away from the critical point, as explained in Appendix
\ref{sec:dev-red}. The BCRG equation follows
Eq.~(\ref{eq: generic_RG_eq}) with $b>1$, while the critical point
and fixed point are identified from the direction of the RG flow and
Eq.~(\ref{eq:Mc_Mf_definition}). 

Applying Eq.\ (\ref{eq: generic_RG_eq}) to the honeycomb model,
choosing $b>1$, yields the BCRG flow shown in
Fig.\ \ref{fig:honeycomb_checkp} (a) that signals the existence of a quantum critical point in the thermodynamic limit.
The ring shape of the Berry curvature
implies two scales in the fBZ that represent the radius and the width of
the ring. They are extracted in the following manner.
Along a particular high symmetry line,
for instance $\widehat{\bm{\phi}}^{\,}_{1}$, the extremum
of the Berry curvature and its location are denoted by
$F^{\,}_{\star,\mathrm{max}}$ and $\phi^{\,}_{1,\mathrm{max}}$,
respectively. Defining the half-extremum as
$\left[
F^{\,}_{\star,\mathrm{max}}+F^{\,}_{\star}(\bm{\phi}^{\,}_{\mathrm{c}},V^{\,}_{2})
\right]/2$, the half-distance between the two
$\phi^{\,}_{1}$'s at which the half-maximum
occurs is denoted by $\phi^{\,}_{1,\mathrm{wid}}$,
as shown schematically in Fig.\ \ref{fig:honeycomb_checkp} (b).
In Appendix~\ref{sec:Wannier_correlation_fn},
we demonstrate that $\xi^{\,}_{1,\mathrm{max}}\:=1/\phi^{\,}_{1,\mathrm{max}}$
and $\xi^{\,}_{1,\mathrm{wid}}\:=1/\phi^{\,}_{1,\mathrm{wid}}$
represent the two length scales over which a correlation function
oscillates and decays, respectively.
Figures~\ref{fig:honeycomb_checkp}~(c) and~(d)
provide supporting evidence for the scaling behaviors
$F^{\,}_{\star,\mathrm{max}}\propto|V^{\,}_{2}-V^{\,}_{2\mathrm{c}}|^{-1}$ and
$\xi^{\,}_{1,\mathrm{max}}\propto\xi^{\,}_{1,\mathrm{wid}}\propto
|V^{\,}_{2}-V^{\,}_{2\mathrm{c}}|^{-1/2}$.
The corresponding scaling exponents
are in full agreement with those of
non-interacting double Weyl nodes\cite{Chen2017}.

\begin{figure}
\includegraphics[clip=true,width=0.99\columnwidth]{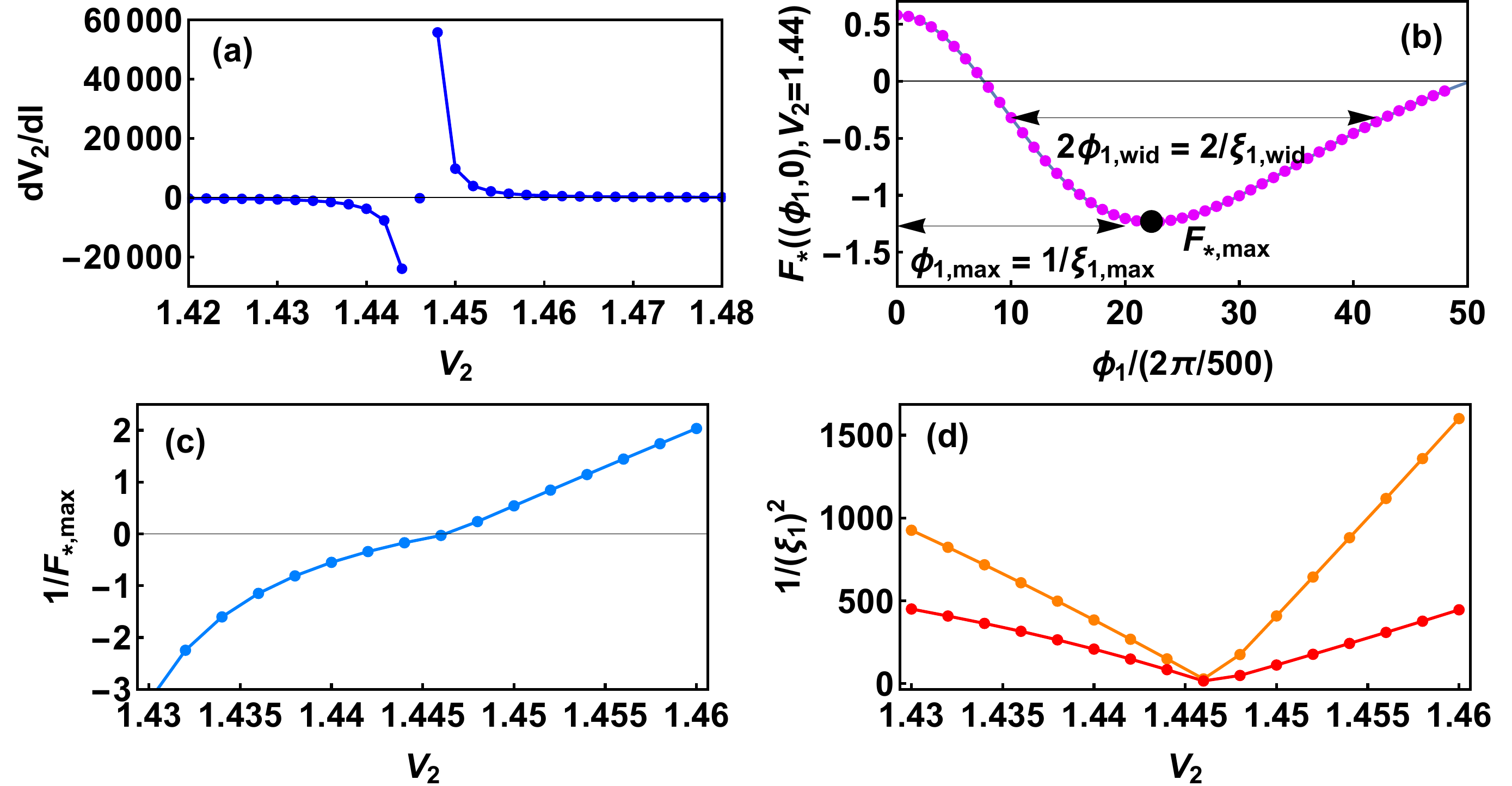}
\caption{
(color online)
(a) BCRG flow of the tuning parameter $V^{\,}_{2}$ for the
honeycomb lattice model, obtained using Eq.\ (\ref{eq: generic_RG_eq})
with the choice $b=2$, the grid spacing $\Delta\bm{\phi}=(3\pi/250,0)$, and
$\Delta V^{\,}_{2}=0.002$. Plotted in units of $\Delta V^{\,}_{2}/\Delta\phi^{2}$,
the BCRG flow signals the existence of 
a quantum critical point at
$V^{\,}_{2\mathrm{c}}\approx1.446$
in the thermodynamic limit.
(b) Schematics of extracting the two correlation lengths
$\xi^{\,}_{1,{\mathrm{max}}}$
and
$\xi^{\,}_{1,{\mathrm{wid}}}$
from the Berry curvature (purple) along the
$\widehat{\bm{\phi}}^{\,}_{1}$ direction.
(c) The inverse 
$1/F^{\,}_{\star,{\mathrm{max}}}$
of the Berry curvature extremum
$F^{\,}_{\star,{\mathrm{max}}}$
versus $V^{\,}_{2}$. The values at $V^{\,}_{2}>V^{\,}_{2\mathrm{c}}$
are enlarged 200 times for readability. The linear behavior near
$V^{\,}_{2\mathrm{c}}$ corresponds to the exponent $\alpha\approx 1$.
(d) Inverses $1/\xi^{2}_{1,\mathrm{max}}$ and $1/\xi^{2}_{1,\mathrm{wid}}$
of the squared correlation lengths
$\xi^{2}_{1,\mathrm{max}}$ (orange) and $\xi^{2}_{1,\mathrm{wid}}$ (red)
versus $V^{\,}_{2}$ plotted in units of $(\pi/500)^{2}$.
The $1/\xi^{2}_{1,\mathrm{wid}}$ at $V^{\,}_{2}>V^{\,}_{2\mathrm{c}}$
is enlarged by 10 times for readability.
The linear behavior near $V_{2\mathrm{c}}$
corresponds to the critical exponent $\nu\approx 1/2$.
        }
\label{fig:honeycomb_checkp}
\end{figure}

\section{Summary and conclusion}
\label{sec: Summary and conclusion}

We have investigated precursor signs
of quantum phase transitions taking place in the thermodynamic limit
between topologically trivial and nontrivial correlated phases by performing
detailed numerical studies of interacting two-dimensional fermionic
and bosonic models harboring FCI states with twisted boundary
conditions. The main result of this work is the observation that,
when inversion or other point-group symmetries are present, the
many-body Berry curvature develops divergences at one or more HSPs in
the space of twists in the boundary conditions, i.e.,
the Brillouin zone for twisted boundary conditions (fBZ),
for critical values of the parameters in our models
defined on finite lattices.
Concomitantly, we find nodal points in the dispersion of
many-body energy levels in the fBZ at the same HSPs. We observe
a connection between the number and dispersion of these nodal points
and the many-body Chern number characterizing the states that partake
in each transition. The many-body energy levels close to the nodal points can
be classified heuristically by the power law that determines the
dispersion in $\bm\phi$-space, in analogy to the topological charge of
Weyl nodes for the non-interacting Bloch bands of semi-metals.
We have determined the exponents $p^{\,}_{d}$
of the dispersion around the nodal points encountered at a
number of topological-to-trivial phase transitions in models harboring
FCI states. At all these points, $p^{\,}_{d}$ is an \textit{integer}.
Finally, we have exploited the above observations to
develop a scaling approach for the many-body Berry curvature around
the HSPs in the fBZ where it becomes divergent. This procedure is shown to
give precursor signs of the thermodynamic phase boundaries of topological phase transitions driven
by either interaction or single-particle parameters, and we have
empirically found different classes of ``criticality'', depending
again on the exponents $p^{\,}_{d}$.

Even though the validity of our methods and observations has been
verified through a battery of rigorous numerical tests for a number of
models, this work is predominantly phenomenological, in the sense that
we do not attempt to prove a number of points, in particular (i) whether the ansatz for the scaling of the Berry curvature is unique, and (ii) what the importance of our findings is in the
thermodynamic limit. We hope to clarify (at least a subset of) the
above points in future work.
Finally, our approach can be generalized to the spin quantum Hall effect,
see Ref.~\onlinecite{Zeng17}.

\section*{Acknowledgements}

S.~K.~is grateful to G.~Palumbo for stimulating discussions. S.~K.~was partially supported through the Boston University Center for Non-Equilibrium Systems and Computation.  M.~S., W.~C., and T.~N.~acknowledge financial support from a grant of the Swiss National Science Foundation.

\appendix

\section{The deviation-reduction mechanism}
\label{sec:dev-red}

The scaling hypothesis takes the differential form captured by Eq.\
(\ref{eq: generic_RG_eq})
for the Berry curvature
$F^{\,}_{\star}(\bm{\phi},\bm{M})$.
According to this hypothesis, 
a flow away from the critical point takes place.
We are going to derive a different scaling equation upon approaching the
fixed-point Berry curvature
$F^{(\mathrm{f})}_{\star}(\bm{\phi},\bm{M}^{\,}_{\mathrm{f}})$
that is defined by demanding that it carries the Chern number
\begin{equation}
C^{\,}_{\star}=
\int\limits_{0}^{2\pi}
\int\limits_{0}^{2\pi}
\frac{\mathrm{d}\phi^{\,}_{1}\,\mathrm{d}\phi^{\,}_{2}}{(2\pi)^{2}}\,
F^{(\mathrm{f})}_{\star}(\bm{\phi},\bm{M}^{\,}_{\mathrm{f}}),
\end{equation}
while it obeys the scaling form
[compare with Eq.\ (\ref{eq: scaling relation})]
\begin{equation}
F^{(\mathrm{f})}_{\star}(\bm{\phi}^{\,}_{\mathrm{c}}
+\delta\bm{\phi},\bm{M}^{\,}_{\mathrm{f}})=
F^{(\mathrm{f})}_{\star}(\bm{\phi}+\delta\bm{\phi}',\bm{M}^{\,}_{\mathrm{f}})
\end{equation}
for some $\delta\bm{\phi}'$ about the HSP $\bm{\phi}^{\,}_{\mathrm{c}}$ in the fBZ.
We make the additive decomposition
\begin{subequations}
\label{eq: additive decomposition}
\begin{equation}
F^{\,}_{\star}(\bm{\phi},\bm{M})=
F^{(\mathrm{f})}_{\star}(\bm{\phi},\bm{M}^{\,}_{\mathrm{f}})
+
\delta F^{(\mathrm{f})}_{\star}(\bm{\phi},\bm{M}),
\end{equation}
where
\begin{equation}
\begin{split}
\delta F^{(\mathrm{f})}_{\star}(\bm{\phi},\bm{M})=&\,
\sum_{\bm{m}\in\mathbb{Z}^{2},|{\bm m}|>0}
\delta F^{(\mathrm{f})}_{\star}(\bm{m},\bm{M})
\\
&\,\times
\cos(m^{\,}_{1}\,\phi^{\,}_{1})\,
\cos(m^{\,}_{2}\,\phi^{\,}_{2})
\end{split}
\end{equation}
so that 
\begin{equation}
0=
\int\limits_{0}^{2\pi}
\int\limits_{0}^{2\pi}
\frac{\mathrm{d}\phi^{\,}_{1}\,\mathrm{d}\phi^{\,}_{2}}{(2\pi)^{2}}\,
\delta F^{(\mathrm{f})}_{\star}(\bm{\phi},\bm{M}),
\end{equation}
while we assume the scaling relation
\begin{equation}
\delta F^{(\mathrm{f})}_{\star}(\bm{\phi}^{\,}_{\mathrm{c}}+\delta\bm{\phi},\bm{M})=
\delta F^{(\mathrm{f})}_{\star}(\bm{\phi}^{\,}_{\mathrm{c}}+\delta\bm{\phi}',\bm{M}')
\end{equation}
\end{subequations}
for some pair $\delta\bm{\phi}'$ and $\bm{M}'$.
If we choose
\begin{equation}
\bm{\phi}^{\,}_{\mathrm{c}}=\bm{0}
\end{equation}
and assume the linear relation
\begin{equation}
\delta\bm{\phi}'=b\,\delta\bm{\phi}
\end{equation}
for the real number $0\leq b$,
we then find the relation
\begin{widetext}
\begin{equation}
\delta F^{(\mathrm{f})}_{\star}(\bm{\phi}^{\,}_{\mathrm{c}}+\delta\bm{\phi}',\bm{M}')
-
\delta F^{(\mathrm{f})}_{\star}(\bm{\phi}^{\,}_{\mathrm{c}}+\delta\bm{\phi}',\bm{M})=
\frac{1}{2}\,
\left(
\delta\bm{\phi}
-
b\,\delta\bm{\phi}'
\right)
\cdot
\left(
\frac{\partial}{\partial\bm{\phi}}
\delta F^{(\mathrm{f})}_{\star}(\bm{\phi}^{\,}_{\mathrm{c}}+\delta\bm{\phi},\bm{M})
\right)
\label{eq:deviation_reduction}
\end{equation}
\end{widetext}
to leading order in an expansion in powers of $\delta\bm{\phi}$.
Contrary to Eq.\ (\ref{eq: generic_RG_eq}),
this equation is of first order in the derivative with respect to
the twisting angles. 

The change of
$\delta F^{(\mathrm{f})}_{\star}$
as $\bm{M}$ changes to $\bm{M}'$, holding
$\bm{\phi}^{\,}_{\mathrm{c}}+\delta\bm{\phi}'$
fixed in the fBZ, is proportional to
the derivative of
$\delta F^{(\mathrm{f})}_{\star}$
with respect to the twisting angle
at $\bm{\phi}^{\,}_{\mathrm{c}}+\delta\bm{\phi}$ in the fBZ
holding $\bm{M}$ fixed. Now,
the decomposition (\ref{eq: additive decomposition})
implies that $\delta F^{(\mathrm{f})}_{\star}$
carries the singularity encoded in
$F^{\,}_{\star}$
upon crossing $\bm{M}^{\,}_{\mathrm{c}}$.
This observation has the following consequences.

For the peak-divergence scenario (in which case $0\leq b<1$),
as one can deduce from Fig.~\ref{fig:correlation_fn}~(a),
when the extremum
$F^{\,}_{\star}(\bm{\phi}^{\,}_{\mathrm{c}},\bm{M})$
is a maximum, then the derivative of
$\delta F^{(\mathrm{f})}_{\star}$ is negative whereas
the multiplicative prefactor $(1-b^{2})/2>0$ is positive.
If so, the left-hand side of Eq.~(\ref{eq:deviation_reduction})
is negative and the Berry curvature divergence is reduced
under the mapping $\bm{M}\rightarrow{\bm M}^{\prime}$.

For the ring-divergence scenario (in which case  $1<b$),
when the extremum
$F^{\,}_{\star}(\bm{\phi}^{\,}_{\mathrm{c}},\bm{M})$
is captured by a double-Lorentzian with a maximum as
shown in Fig.~\ref{fig:correlation_fn} (b), then the derivative of
$\delta F^{(\mathrm{f})}_{\star}$ is positive whereas
the multiplicative prefactor $(1-b^{2})/2<0$ is negative.
If so, the left-hand side of Eq.~(\ref{eq:deviation_reduction})
is negative and the Berry curvature divergence is also reduced
under the mapping $\bm{M}\rightarrow{\bm M}^{\prime}$.

If the extrema of many-body Berry curvature are minima instead of maxima,
the same logic leads to a positive
left-hand side of Eq.~(\ref{eq:deviation_reduction}),
so the negative divergence of Berry curvature is again reduced.

We conclude that the divergence of the many-body Berry curvature
is reduced under this scaling procedure.
The system is gradually flowing away from the value $\bm{M}^{\,}_{\mathrm{c}}$
at which the Berry curvature diverges.

\section{A scaling law in the peak-divergence scenario}
\label{appen:scaling_law}

In the peak-divergence scenario of Sec.~\ref{sec:peak_divergence}, it
is instructive to calculate the contribution
\begin{equation}
\begin{split}
C^{\,}_{\star\,\mathrm{div}}(\bm{M})\:=&\,
\int\limits_{-\xi^{-1}_{1}}^{+\xi^{-1}_{1}}
\frac{\mathrm{d}\,\delta\phi^{\,}_{1}}{2\pi}
\int\limits_{-\xi_{2}^{-1}}^{+\xi_{2}^{-1}}
\frac{\mathrm{d}\,\delta\phi^{\,}_{2}}{2\pi}\,
\\
&\,
\times
\frac{
F^{\,}_{\star}(\bm{\phi}^{\,}_{\mathrm{c}},\bm{M})
     }
     {
1
+
\xi^{2}_{1}(\bm{M})\,\delta\phi^{2}_{1}
+
\xi^{2}_{2}(\bm{M})\,\delta\phi^{2}_{2}
     }
\end{split}
\end{equation}
to the many-body Chern number that arises from the proximity of
$\bm{\phi}^{\,}_{\mathrm{c}}+\delta\bm{\phi}$ to
$\bm{\phi}^{\,}_{\mathrm{c}}$, as approximated by the OZ ansatz.
Here, we assume that
$F^{\,}_{\star}(\bm{\phi}^{\,}_{\mathrm{c}},\bm{M})$ and
$\xi^{\,}_{i}(\bm{M})$, with $i=1,2$, obey the singular scalings
(\ref{eq: scaling ansatz Ornstein-Zernike a}) and
(\ref{eq: scaling ansatz Ornstein-Zernike b}), respectively.
It is
\begin{align}
C^{\,}_{\star\,\mathrm{div}}(\bm{M})=
\mathrm{const}\times
\frac{
F^{\,}_{\star}(\bm{\phi}^{\,}_{\mathrm{c}},\bm{M})
     }
     {
\xi^{\,}_{1}(\bm{M})\,\xi^{\,}_{2}(\bm{M})
     }.
\end{align}
Since the left-hand side is always finite, although not necessarily constant, 
on either side of $\bm{M}^{\,}_{\mathrm{c}}$,
the divergence of
$\lim_{\bm{M}\to\bm{M}^{\,}_{\mathrm{c}}}F^{\,}_{\star}({\bm{\phi}}_{\mathrm{c}},\bm{M})$
must the be compensated by the divergences of
$\lim_{\bm{M}\to\bm{M}^{\,}_{\mathrm{c}}}\xi^{\,}_{i}(\bm{M})$.
By matching the divergences of the numerator and denominator
on the right-hand side, we deduce the Ornstein-Zernike scaling law
\begin{equation}
\alpha\approx\nu^{\,}_{1}+\nu^{\,}_{2}.
\end{equation}

\section{Wannier state correlation function}
\label{sec:Wannier_correlation_fn}

Our starting point are the twisted boundary conditions that span the fBZ
\begin{equation}
\left\{
(\phi^{\,}_{1},\phi^{\,}_{2})\,|\,0\leq\phi^{\,}_{i}<2\pi,
\quad i=1,2
\right\}.
\end{equation}
We do the rescaling of the fBZ defined by
\begin{equation}
\varphi^{\,}_{i}\:=\frac{1}{L^{\,}_{i}\,|\bm{a}^{\,}_{i}|}\phi^{\,}_{i},
\qquad i=1,2,
\end{equation}
where $\bm{a}^{\,}_{i}$ are the basis vector of the finite lattices made of
\begin{equation}
L^{\,}_{1}\times L^{\,}_{2}
\end{equation}
repeated unit cells that were considered
in Secs.\ \ref{sec: Berry curvature and Chern number ...} and
\ref{sec: FCI-to-trivial transitions}.
The rescaled fBZ has the area
\begin{equation}
\prod_{i=1}^{2}\frac{2\pi}{L^{\,}_{i}\,|\bm{a}^{\,}_{i}|}
\end{equation}
and $\varphi^{\,}_{i}$ has the units of inverse length.
We interpret the rescaled fBZ as the Brillouin zone of
an auxiliary dual lattice of infinite extent
spanned by the lattice vectors
\begin{equation}
\bm{R}^{\,}_{\bm{m}}\:=
\sum_{i=1}^{2}
m^{\,}_{i}\,L^{\,}_{i}\,\bm{a}^{\,}_{i},
\qquad
\bm{m}\equiv(m^{\,}_{1},m^{\,}_{2})\in\mathbb{Z}^{2}.
\label{eq: def auxiliary lattice}
\end{equation}
We can associate the Fock space of the many-body Hamiltonians $\widehat{H}$
defined in Sec.\ \ref{sec: FCI-to-trivial transitions}
to each site $\bm{R}^{\,}_{\bm{m}}$.
Any one of the  Hamiltonians $\widehat{H}$
considered in Sec.\ \ref{sec: FCI-to-trivial transitions}
has a GS $|\Psi^{\,}_{\star}(\bm{\phi},\bm{M})\rangle$
with the wave function in position space
\begin{equation}
\Psi^{\,}_{\star}(\bm{r}^{\,}_{1},\cdots,\bm{r}^{\,}_{N}|\bm{\phi},\bm{M})\:=
\langle\bm{r}^{\,}_{1},\cdots,\bm{r}^{\,}_{N}
|\Psi^{\,}_{\star}(\bm{\phi},\bm{M})\rangle.
\label{eq: def Psi star R's phi's M's}
\end{equation}
By assumption, this GS carries a non-vanishing many-body Chern
number
\begin{equation}
C^{\,}_{\star}(\bm{M})\:=
\int\limits_{0}^{2\pi}\int\limits_{0}^{2\pi}
\frac{\mathrm{d}\phi^{\,}_{1}\,\mathrm{d}\phi^{\,}_{2}}{(2\pi)^{2}}
F^{\,}_{\star}(\bm{\phi},\bm{M}),
\end{equation}
where the Berry curvature
$F^{\,}_{\star}(\bm{\phi},\bm{M})$
was defined in Eq.\ (\ref{eq:berry-alt})
in some region of parameter space $\{\bm{M}\in\mathbb{R}^{m}\}$.
For notational simplicity, we shall denote with
\begin{equation}
\Psi^{\,}_{\star}(\bm{r}^{\,}_{1},\cdots,\bm{r}^{\,}_{N}|\bm{\varphi},\bm{M})\:=
\langle\bm{r}^{\,}_{1},\cdots,\bm{r}^{\,}_{N}
|\Psi^{\,}_{\star}(\bm{\varphi},\bm{M})\rangle
\label{eq: def Psi star R's varphi's M's}
\end{equation}
the function obtained by expressing $\bm{\phi}$ in terms of
$\bm{\varphi}$ in Eq.\ (\ref{eq: def Psi star R's phi's M's}).
With the help of Eq.\ (\ref{eq:twisted_bc}),
we have the relation
\begin{equation}
\begin{split}
&
\Psi^{\,}_{\star}
(\bm{r}^{\,}_{1}+\bm{R}^{\,}_{\bm{m}},\cdots,\bm{r}^{\,}_{N}+\bm{R}^{\,}_{\bm{m}}|
\bm{\varphi},\bm{M})=
\\
&
\qquad\qquad\qquad\qquad
e^{\mathrm{i}N\,\bm{\varphi}\cdot\bm{R}^{\,}_{\bm{m}}}\,
\Psi^{\,}_{\star}(\bm{r}^{\,}_{1},\cdots,\bm{r}^{\,}_{N}|\bm{\varphi},\bm{M})
\end{split}
\end{equation}
for any site $\bm{R}^{\,}_{\bm{m}}$ of the lattice dual to the rescaled fBZ that
was defined in Eq.\ (\ref{eq: def auxiliary lattice}).
If we define
\begin{equation}
\bm{X}^{\,}_{\bm{m}}\:=
N\,\bm{R}^{\,}_{\bm{m}},  
\end{equation}
we recognize that
\begin{equation}
\begin{split}
&
\Psi^{\,}_{\star}
(\bm{r}^{\,}_{1}+N^{-1}\,\bm{X}^{\,}_{\bm{m}},\cdots,
\bm{r}^{\,}_{N}+N^{-1}\,\bm{X}^{\,}_{\bm{m}}|
\bm{\varphi},\bm{M})=
\\
&
\qquad\qquad\qquad\qquad
e^{\mathrm{i}\bm{\varphi}\cdot\bm{X}^{\,}_{\bm{m}}}\,
\Psi^{\,}_{\star}(\bm{r}^{\,}_{1},\cdots,\bm{r}^{\,}_{N}|\bm{\varphi},\bm{M})
\end{split}
\label{eq: Bloch thm for Psi star}
\end{equation}
takes the form of Bloch's theorem, whereby the center of mass
\begin{equation}
\bm{X}\:=\sum_{i=1}^{N}\bm{r}^{\,}_{i}
\end{equation}
of the $N$ quantum particles has been translated by $\bm{X}^{\,}_{\bm{m}}$.
Correspondingly, we define the center of mass operator
\begin{equation}
\widehat{\bm{X}}\:=
\sum_{i=1}^{N}\widehat{\bm{r}}^{\,}_{i}
\end{equation}
together with the Wannier state supported on the lattice dual to the
rescaled fBZ that is defined by
\begin{widetext}
\begin{equation}
|W^{\,}_{\star}(\bm{X}^{\,}_{\bm{m}},\bm{M})\rangle\:=
\left(
\prod_{i=1}^{2}
\int\limits_{0}^{2\pi/(L^{\,}_{i}\,|\bm{a}^{\,}_{i}|)}
\frac{\mathrm{d}\varphi^{\,}_{i}}{2\pi/(L^{\,}_{i}\,|\bm{a}^{\,}_{i}|)}
\right)\,
e^{+\mathrm{i}\bm{\varphi}\cdot\left(\widehat{\bm{X}}-\bm{X}^{\,}_{\bm{m}}\right)}\,
|\Psi^{\,}_{\star}(\bm{\varphi},\bm{M})\rangle.
\label{eq: def Wannier in terms Bloch states}
\end{equation}
Inversion of Eq.\ (\ref{eq: def Wannier in terms Bloch states})
gives
\begin{equation}
|\Psi^{\,}_{\star}(\bm{\varphi},\bm{M})\rangle=
\sum_{\bm{m}\in\mathbb{Z}^{2}}
e^{-\mathrm{i}\bm{\varphi}\cdot\left(\widehat{\bm{X}}-\bm{X}^{\,}_{\bm{m}}\right)}\,
|W^{\,}_{\star}(\bm{X}^{\,}_{\bm{m}},\bm{M})\rangle.
\end{equation}
By construction, the Wannier wave function is defined to be
\begin{equation}
\begin{split}
W^{\,}_{\star}
(\bm{r}^{\,}_{1},\cdots,\bm{r}^{\,}_{N}|\bm{X}^{\,}_{\bm{m}},\bm{M})\:=&\,
\langle\bm{r}^{\,}_{1},\cdots,\bm{r}^{\,}_{N}
|W^{\,}_{\star}(\bm{X}^{\,}_{\bm{m}},\bm{M})\rangle
\\
=&\,
\left(
\prod_{i=1}^{2}
\int\limits_{0}^{2\pi/(L^{\,}_{i}\,|\bm{a}^{\,}_{i}|)}
\frac{\mathrm{d}\varphi^{\,}_{i}}{2\pi/(L^{\,}_{i}\,|\bm{a}^{\,}_{i}|)}
\right)\,
e^{+\mathrm{i}\bm{\varphi}\cdot\left(\bm{X}-\bm{X}^{\,}_{\bm{m}}\right)}\,
\Psi^{\,}_{\star}(\bm{r}^{\,}_{1},\cdots,\bm{r}^{\,}_{N}|\bm{\varphi},\bm{M}).
\end{split}
\end{equation}
The expectation value of the center of mass operator $\widehat{\bm{X}}$
in the Wannier state  $|W^{\,}_{\star}(\bm{X}^{\,}_{\bm{m}},\bm{M})\rangle$
is as localized as may be about $\bm{X}^{\,}_{\bm{m}}$
in the lattice dual to the rescaled fBZ.
The Fourier transform of the many-body Berry curvature defined by
Eq.\ (\ref{eq:berry-alt}) is denoted
\begin{equation}
F^{\,}_{\star\,\bm{X}^{\,}_{\bm{m}}}(\bm{M})\:= 
\left(
\prod_{i=1}^{2}
\int\limits_{0}^{2\pi/(L^{\,}_{i}\,|\bm{a}^{\,}_{i}|)}
\frac{\mathrm{d}\varphi^{\,}_{i}}{2\pi/(L^{\,}_{i}\,|\bm{a}^{\,}_{i}|)}
\right)\,
e^{+\mathrm{i}\bm{\varphi}\cdot\bm{X}^{\,}_{\bm{m}}}\,
F^{\,}_{\star}(\bm{\varphi},\bm{M}).
\end{equation} 
It is represented in terms of the Wannier state by
the correlation function
\cite{Marzari2012,Gradhand2012,Wang2006}
\begin{equation}
F^{\,}_{\star\,\bm{X}^{\,}_{\bm{m}}}(\bm{M})= 
-\mathrm{i}
\langle W^{\,}_{\star}(\bm{X}^{\,}_{\bm{m}},\bm{M})|
\left(
X^{\,}_{\bm{m}\,1}\,
\widehat{X}^{\,}_{2}
-
X^{\,}_{\bm{m}\,2}\,
\widehat{X}^{\,}_{1}
\right)
|W^{\,}_{\star}(\bm{0},\bm{M})\rangle.
\label{eq:Wannier_correlation}
\end{equation}
\end{widetext}

If we use the OZ ansatz
\begin{equation}
F^{\,}_{\star}(\bm{\varphi}^{\,}_{\mathrm{c}}+\delta\bm{\varphi},\bm{M})\sim
\frac{
F^{\,}_{\star}(\bm{\varphi}^{\,}_{\mathrm{c}},\bm{M})
     }
     {
1
+
\sum_{i=1}^{2}
[\xi^{\,}_{i}(\bm{M})\,L^{\,}_{i}\,|\bm{a}^{\,}_{i}|]^{2}\,\delta\varphi^{2}_{i}
     }
\label{eq:Ornstein-Zernike bis}
\end{equation}
to fit the peak-divergence scenario,
we then identify the characteristic width
$[\xi^{\,}_{i}(\bm{M})\,L^{\,}_{i}\,|\bm{a}^{\,}_{i}|]^{-1}$ of
$F^{\,}_{\star}(\bm{\varphi},\bm{M})$
as a function of $\bm{\varphi}$
in the rescaled fBZ.
Upon Fourier transform of the OZ ansatz
(\ref{eq:Ornstein-Zernike bis}), the correlation function
$F^{\,}_{\star\,\bm{X}^{\,}_{\bm{m}}}(\bm{M})$ has the
characteristic decay length in the lattice dual to the rescaled fBZ given by
$\xi^{\,}_{i}(\bm{M})\,L^{\,}_{i}\,|\bm{a}^{\,}_{i}|$. 
In the ring-divergence
scenario, the Fourier transform of the
ring-shaped Berry curvature in Fig.~\ref{fig:honeycomb_checkp}(b)
gives a correlation function
$F^{\,}_{\star\,\bm{X}^{\,}_{\bm{m}}}(\bm{M})$ that oscillates with
$\xi^{\,}_{\mathrm{max}}(\bm{M})\,L^{\,}_{i}\,|\bm{a}^{\,}_{i}|$
and decays with
$\xi^{\,}_{\mathrm{wid}}(\bm{M})\,L^{\,}_{i}\,|\bm{a}^{\,}_{i}|$. 

This Wannier-state formalism provides an appealing interpretation for
the singularities of the Berry curvature in the fBZ for finite systems. As
detailed in Sec.\ \ref{sec:Berry_curvature_Chern_number}, the choice
of twisted boundary conditions in the continuous range
$[0,2\pi)\times [0,2\pi)$
forms an auxiliary space that allows for the
definition of the Chern number. In the thermodynamic limit,
this Chern number becomes the quantized Hall conductivity. At the same time,
imposing these twisted boundary conditions on many-body wavefunctions as in
Eq.\ (\ref{eq: Bloch thm for Psi star})
allows for a definition of an infinite lattice \textit{dual} to the fBZ.
On this infinite lattice, which should not be confused with
the original finite lattice $\Lambda$ on which the many-body Hamiltonians have
been diagonalized, the functions
$\xi^{\,}_{i}(\bm{M})\,L^{\,}_{i}\,|\bm{a}^{\,}_{i}|$ with $i=1,2$,
take the meaning of correlation lengths. Moreover,
the Fourier transform over the rescaled fBZ of the Berry curvature
$F^{\,}_{\star}(\bm{\varphi},\bm{M})$
that delivers $F^{\,}_{\star\,\bm{X}^{\,}_{\bm{m}}}(\bm{M})$
can be interpreted as a
a correlation function that measures the overlap of
Wannier states centered at two copies of the finite lattice $\Lambda$ 
that are distance $\bm{X}^{\,}_{\bm{m}}$
away on this dual lattice. It is in this
respect that the many-body Berry curvature
for any exact eigenstate of the many-body 
Hamiltonian defined on the \textit{finite} lattice $\Lambda$
is similar to scaling functions in critical phenomena.
The same is true when identifying the divergence of the
$\xi(\bm{M})$'s at $\bm{M}^{\,}_{\mathrm{c}}$ and
their vanishing at $\bm{M}^{\,}_{\mathrm{f}}$ with the
notion of scale invariance.

For non-interacting tight-binding 
Hamiltonians for which Bloch's theorem holds upon imposing
periodic boundary conditions on a finite lattice,
using twisted boundary conditions allows to explore the
Brillouin zone in the thermodynamic limit. It is for this reason
that the critical behavior of the Berry curvature
for the finite number of occupied states in the valence band
gives access to some thermodynamic critical exponents
\cite{Chen2016,Chen2016b,Chen2017}.

For interacting Hamiltonians defined on a finite lattice $\Lambda$,
the critical behavior of the Berry curvature
needs to be related to the quantum phase transition in the
thermodynamic limit. To prove or disprove such a relation, it is
necessary to perform finite-size scaling numerically, which is not
possible with any numerical method available today.
A diverging correlation length corresponds to a
critical finite-size Wannier many-body state.
A large correlation length is interpreted as 
the finite-size interacting system being close to a gap-closing
many-body level crossing.
A short correlation length implies
the finite-size interacting system is deep inside
a gapped (trivial or nontrivial) topological phase.
This is irrespective of whether the quantum
phase transition involves a change in a Landau order parameter or not.

\section{Symmetries of many-body Hamiltonians in flux space}
\label{sec:Symm}

In this appendix we show how spatial symmetries of an interacting
tight-binding Hamiltonian $\widehat{H}(\bm{0})$, present when the
(finitely sized) system is defined on a lattice with periodic boundary
conditions, translate into symmetries of $\widehat{H}(\bm{\phi})$ in
flux space when twisted boundary conditions $\bm{\phi}$ are imposed.

Consider a symmetry operation $\widehat{S}$ and a tight-binding
Hamiltonian with a density-density interaction
\begin{equation}
\begin{split}
\widehat{H}(\bm{\phi})=&\,
\sum_{\bm{k}\in\mathrm{BZ}}
\sum_{\alpha,\beta}
\widehat{c}^{\dag}_{\bm{k},\alpha}\,
\mathcal{H}^{\,}_{\bm{k}+\bm{\phi},\alpha,\beta}\,
\widehat{c}^{\,}_{\bm{k},\beta}
\\
&\,+
\sum_{\bm{i},\bm{j};\alpha,\beta}
V^{\,}_{\bm{i},\bm{j};\alpha,\beta}\,
\widehat{n}^{\,}_{\bm{i},\alpha}\,
\widehat{n}^{\,}_{\bm{j},\beta},
\end{split}
\end{equation}
where $\alpha$ and $\beta$ label local degrees of freedom in the unit
cell such as orbitals and spin.  Due to its density-density form, the
interaction term is unaffected by the boundary conditions, which only
affect the quadratic part
$\mathcal{H}^{\,}_{\bm{k}+\bm{\phi},\alpha,\beta}$.

The action of $\widehat{S}$ on the second quantized operators can be
represented by a unitary $B^{\,}_{\bm{k};\alpha,\beta}$
\begin{subequations}
\begin{align}
&
\widehat{S}\,
\widehat{c}^{\dag}_{\bm{k},\alpha}\,
\widehat{S}^{-1}=
\sum_{\beta}
B^{\,}_{\bm{k};\alpha,\beta}\,
\widehat{c}^{\dag}_{S(\bm{k}),\beta},
\\
&
\left(
\widehat{S}\,
\widehat{c}^{\dag}_{\bm{k},\alpha}\,
\widehat{S}^{-1}
\right)^{\dag}=
\sum_{\beta}
B^{*}_{\bm{k};\alpha,\beta}\,
\widehat{c}^{\,}_{S(\bm{k}),\beta},
\end{align}
\end{subequations}
where $S(\bm{k})$ applies the spatial transformation (like a
reflection or rotation) to the momentum.  By assumption, the system
with periodic boundary conditions is invariant under $\widehat{S}$,
i.e.,
\begin{equation}
\widehat{S}\,
\widehat{H}(\bm{0})\,
\widehat{S}^{-1}=
\widehat{H}(\bm{0}).
\end{equation}
If we assume that $\widehat{S}^{-1}=\widehat{S}^{\dag}$,
this implies
\begin{equation}
\mathcal{H}^{\,}_{S(\bm{k}),\alpha,\beta}=
\sum_{\alpha',\beta'}
B^{\,}_{\bm{k};\alpha^{\prime},\alpha}\,
\mathcal{H}^{\,}_{\bm{k},\alpha',\beta'}\,
B^{*}_{\bm{k};\beta^{\prime},\beta}.
\end{equation}
Using this relation, one shows that
\begin{equation}
\widehat{S}\,
\widehat{H}(\bm{\phi})\,
\widehat{S}^{-1}=
\widehat{H}(S(\bm{\phi}))
\end{equation}
Spatial symmetries thus imply that the Hamiltonians with boundary
conditions $\bm{\phi}$ and $S(\bm{\phi})$ are unitarily related. It
follows that gauge-invariant quantities like the many-body Berry
curvature obey the same symmetries, even in finite-size systems
\begin{equation}
F^{\,}_{n}( \bm{\phi},\bm{M})=F^{\,}_{n}( S(\bm{\phi}),\bm{M}).
\end{equation}

\end{document}